\documentclass{article}

\usepackage{booktabs} 
\usepackage{titling}
\usepackage{graphicx}
\usepackage{xpatch}
\usepackage{hyperref}
\usepackage{setspace}
\usepackage{multirow}
\usepackage{tabularray}
\usepackage{tabularx}
\usepackage{colortbl}
\usepackage{color}
\usepackage{array}
\usepackage[most]{tcolorbox}
\usepackage{caption} 
\tcbuselibrary{breakable}
\usepackage{subcaption}
\usepackage{amsmath}
\usepackage{amssymb}
\usepackage{float}
\usepackage{longtable}
\usepackage{epstopdf}
\usepackage{natbib}
\usepackage{makecell} 
\renewcommand{\thefootnote}{\fnsymbol{footnote}}
\usepackage{tikz}
\usetikzlibrary{arrows.meta,positioning,shapes.geometric,fit,backgrounds}

\bibpunct[, ]{(}{)}{;}{a}{}{,}

\newcommand{\code}[1]{\texttt{#1}}

\hypersetup{
	colorlinks,
	linkcolor={orange!95!black},
	citecolor={blue!90!black},
	urlcolor={olive!95!black}
}

\begin{document}

\title{On the Carbon Footprint of Economic Research in the Age of Generative AI}

\author{
  Andres Alonso-Robisco\textsuperscript{$\dagger$}\textsuperscript{$\ddagger$}\thanks{Corresponding author: \texttt{andres.alonso@bde.es}. \\ The views and analyses expressed in this work are the sole responsibility of the authors and do not necessarily reflect those of Banco de Espa\~na or the Eurosystem.}
  , 
  Carlos Esparcia\textsuperscript{$\ddagger$}, 
  and Francisco Jare\~no\textsuperscript{$\ddagger$} \\
  \textsuperscript{$\dagger$}{\footnotesize Banco de Espa\~na}\\
  \textsuperscript{$\ddagger$}{\footnotesize Universidad de Castilla-La Mancha}
}

\date{\today}

\maketitle
\setcounter{footnote}{0}
\renewcommand{\thefootnote}{\arabic{footnote}}

\begin{abstract}
Generative artificial intelligence (AI) is increasingly used to write and refactor research code, expanding computational workflows. At the same time, Green AI research has largely measured the footprint of models rather than the downstream workflows in which GenAI is a tool. We shift the unit of analysis from models to workflows and treat prompts as decision policies that allocate discretion between researcher and system, governing what is executed and when iteration stops. We contribute in two ways. First, we map the recent Green AI literature into seven themes: training footprint is the largest cluster, while inference efficiency and system level optimisation are growing rapidly, alongside measurement protocols, green algorithms, governance, and security and efficiency trade-offs. Second, we benchmark a modern economic survey workflow, an LDA-based literature mapping implemented with GenAI assisted coding and executed in a fixed cloud notebook, measuring runtime and estimated CO2e with \code{CodeCarbon}. Injecting generic green language into prompts has no reliable effect, whereas operational constraints and decision rule prompts deliver large and stable footprint reductions while preserving decision equivalent topic outputs. The results identify human in the loop governance as a practical lever to align GenAI productivity with environmental efficiency.
\end{abstract}

\vspace{0.5em}
\vspace{1em}

\clearpage
\tableofcontents
\newpage

\section{Introduction}\label{sec:intro}

Generative artificial intelligence (GenAI) is rapidly becoming part of the production technology of economic research. A growing literature argues that large language models can support economists across the research cycle, from background search and drafting to data analysis, mathematical reasoning, and coding \citep{korinek2023jel_genai_econ,korinek2023nber_language_models}. Recent work further conceptualises these systems as increasingly agentic research assistants that interact iteratively with data, code, and intermediate outputs, reshaping workflows rather than simply accelerating isolated tasks \citep{korinek2025agents}. Empirical evidence is consistent with a meaningful productivity channel: GenAI adoption correlates with higher research output and moderate quality gains in the social and behavioral sciences \citep{filimonovic2025can}, and with higher scientific impact and earlier transitions into leadership roles at scale in the natural sciences \citep{hao2025_ai_tools_nature}. Taken together, these contributions suggest that GenAI enabled research methods are changing how we produce, scale, and evaluate research.

Coding is a central margin in this transformation. As empirical workflows grow in complexity, researchers increasingly rely on code to implement, update, and audit computational pipelines. This matters especially for economic surveys and evidence synthesis. Surveys remain a core piece of the discipline's knowledge infrastructure, but scale increasingly shapes their production. The growth of working paper series and journal outlets has made purely narrative synthesis harder to sustain, encouraging text as data pipelines that provide a first pass map of a field and guide subsequent close reading \citep{ash2023textalgorithms,gentzkow2019text}.

Within this toolkit, topic models, and in particular Latent Dirichlet Allocation (LDA), have become a pragmatic default for building quantitative taxonomies of large corpora. LDA offers an interpretable representation in which documents load on multiple themes and themes are defined by word distributions \citep{blei2003lda,blei2012probabilistic}. Applications in economics and adjacent fields use topic models in a survey oriented way: to describe field evolution, identify clusters, and discipline what gets read in depth \citep{ambrosino2018topicmodeling,wehrheim2019economichistory}. Mainstream software environments have also made these methods operationally accessible to economists \citep{schwarz2018ldagibbs}. GenAI adds a further layer to this survey pipeline by lowering the time and cognitive cost of writing and iterating the underlying code, including end to end implementations of workflows such as LDA.

At the same time, wider reliance on AI systems has renewed attention to environmental footprint. Early contributions documented the substantial energy requirements associated with training modern machine learning models \citep{strubell2019energy}. Subsequent research shows that, for large language models, inference can account for a large share of lifecycle energy use as models deploy at scale \citep{jegham2025hungry}. Beyond carbon emissions, AI systems also require water and other inputs, raising broader sustainability concerns \citep{li2023_thirsty}. These concerns have motivated the Green AI literature, which emphasises energy efficient algorithms, transparent reporting of computational costs, and evaluation of progress along efficiency dimensions in addition to predictive performance \citep{schwartz2019greenai,verdecchia2023greenai}. Policy initiatives and regulation increasingly incorporate sustainability considerations as well \citep{EU2024_AIACT, OECD2019_AI_Recommendation, UNESCO2021_Ethics_AI,MINECO2022_algoritmos_verdes}.

Despite their shared relevance for modern research practice, the literatures on GenAI in economics and on Green AI remain weakly connected. Work on GenAI in economics focuses on productivity, research quality, and methodological innovation, while it largely abstracts from the environmental implications of AI-assisted work. Conversely, Green AI has concentrated on the footprint of models, especially training and inference, and it pays less attention to downstream research workflows where GenAI acts as a general purpose tool for producing code and running pipelines. As a result, we lack systematic evidence on the environmental footprint of GenAI enabled economic research workflows, and on how that footprint depends on how researchers structure interaction with these systems.

We argue that the missing link is not only empirical but also conceptual. Much of Green AI takes the model as the unit of analysis. In applied research production, the relevant unit often is a workflow: a sequence of coding and execution steps in which a language model can increase iteration, widen exploration, and produce intermediate artefacts that a researcher might not otherwise compute. In that setting, prompt design functions as a decision policy that allocates responsibility between the researcher and the system. This allocation governs what the workflow executes and when it stops, shaping runtime and CO$_2$e alongside any productivity gains from AI-assisted coding.

This paper addresses this gap. We first update and map the fast growing Green AI literature, organising recent contributions into a tractable taxonomy that is directly relevant for research production as GenAI adoption scales. We highlight what is relatively well established versus still fragmented, and we use this synthesis to motivate a workflow-level perspective in which the unit of analysis is the downstream pipeline that researchers actually run, rather than the model in isolation.

Second, we study the energy and emissions implications of AI-assisted coding in a benchmark research production setting. We focus on computational governance: how prompt design structures delegation of decisions such as exploration scope, stopping rules, and which outputs the workflow computes and stores. We operationalise this mechanism in a transparent and replicable way by benchmarking a survey-style literature mapping workflow based on unsupervised text methods, taking Latent Dirichlet Allocation (LDA) as a representative technique that economists and finance researchers use to structure unlabelled corpora \citep{blei2003lda,blei2012probabilistic,gentzkow2019text}.\footnote{The choice of LDA is not central in itself. It provides a transparent and replicable pipeline through which we can observe how different coding choices translate into differences in computational cost.} Notably, our empirical footprint measures cover execution of the AI-assisted notebook workflow (runtime and \texttt{CodeCarbon} estimates), not the data centre footprint of Large Language Models (LLMs) inference used to generate code or filter documents, which requires different instrumentation and is studied in a separate literature \citep{jegham2025hungry,kadian2025tokenpowerbench,lannelongue2021green}.Therefore, we present our benchmark as an illustrative mechanism and a replicable measurement template, not as a global estimate of the footprint of AI-assisted research. Nevertheless, making this part of the footprint measurable and comparable at the workflow level can help researchers manage how they interact with GenAI in practice, by translating prompt and design choices into quantifiable differences in runtime and emissions, which is increasingly important as GenAI becomes embedded in routine research production.

To make the contribution explicit, Table~\ref{tab:research_questions} summarises our research questions, objectives, and motivations.

\begin{table}[H]
\centering
\caption{Research questions, objectives, and motivations.}
\label{tab:research_questions}
\resizebox{0.85\linewidth}{!}{%
\begin{tabularx}{\linewidth}{@{}l X X@{}}
\toprule
RQ & Objective & Motivation \\
\midrule
RQ1 & Quantify how alternative prompting strategies change the computational footprint of a decision-equivalent survey workflow, measured by runtime and CO$_2$e. & GenAI lowers the cost of producing and iterating code, which can increase throughput. The environmental cost of this productivity gain is not well measured at the workflow level, where repeated execution and redundant computation can arise. \\

RQ2 & Test whether generic green language is sufficient to reduce footprint, or whether operational constraints and explicit decision rules are required to deliver stable reductions in runtime and emissions. & In practice, researchers often rely on high level instructions when interacting with LLMs. If such instructions do not reliably change what the system computes, then meaningful footprint reductions require governance through explicit constraints, not only normative appeals. \\

RQ3 & Assess practical output equivalence across prompting strategies by testing whether reduced footprint comes with any substantive change in topic model outputs. & Footprint reductions are only meaningful if the analytical object is preserved. Because topic models are not identified across runs, we need a transparent equivalence check to separate genuine efficiency gains from changes in the underlying clustering of documents. \\

RQ4 & Conceptualise and test prompts as decision policies that allocate discretion between the researcher and the model, shaping the breadth of search and the amount of computation executed in AI-assisted research. & The key margin is not whether an LLM is used, but how human machine interaction is structured. Prompt design can be treated as part of computational governance, since it determines stopping rules, search scope, and which outputs are computed and stored. \\
\bottomrule
\end{tabularx}%
}
\end{table}

The remainder of the paper is organised to foreground the survey contribution and then use the benchmark as a focused complement. Section~\ref{sec:lit_review} links two strands that have largely evolved in parallel: GenAI as a research production technology and Green AI as a literature on measurement and efficiency. Section~\ref{sec:greenai_map} provides the core survey output, mapping the Green AI field into interpretable themes and extracting implications and open questions that matter for applied research workflows. Section~\ref{sec:implementation} then present the controlled prompting benchmark: we operationalise prompts as decision policies and measure how alternative interaction designs translate into differences in runtime and CO$_2$e under a fixed workflow and environment, while checking practical output equivalence. Section~\ref{sec:conclusions} summarises implications for computational governance and proposes minimal reporting practices for computational surveys in the age of GenAI.

\section{Literature review}\label{sec:lit_review}

The rapid diffusion of GenAI has prompted a growing literature on its role in academic research, particularly within economics and related social sciences. Early contributions frame large language models as general-purpose research tools that can assist economists across multiple stages of the research process, including literature review, writing, data analysis, mathematical reasoning, and code generation \citep{korinek2023jel_genai_econ, korinek2023nber_language_models}. Rather than acting as task specific algorithms, these systems increasingly function as interactive and adaptive research assistants, capable of responding to intermediate outputs and shaping research workflows in an iterative manner \citep{korinek2025agents}. Related perspectives in AI for science emphasize the breadth of this integration across disciplines, from hypothesis formation and experimental design to analysis and discovery, including generative methods for scientific content and design \citep{Wang2023_scientific_discovery_ai}.

Complementing this conceptual work, recent empirical studies have begun to quantify the effects of AI adoption on research outcomes and scientific production. Evidence at scale in the natural sciences links AI tool usage to higher individual impact alongside a contraction in the collective breadth of scientific focus \citep{hao2025_ai_tools_nature}. In parallel, measurement work documents that the direct use and potential benefits of AI appear widespread across disciplines and have grown rapidly over the past decade, while highlighting heterogeneity and gaps in how AI capabilities diffuse into research practice \citep{GaoWang2024_AI_science_nathumbe}. Within the social and behavioral sciences, GenAI adoption is associated with higher research productivity and moderate gains in publication quality, with stronger effects for early-career researchers and for tasks involving complex technical or linguistic components \citep{filimonovic2025can}. Together, this evidence supports the view that GenAI can relax constraints in research production, but that realized gains depend on where and how researchers integrate these systems into their workflows.

Within economics, coding occupies a central position in modern empirical work. The expansion of the ``text as data'' paradigm and the widespread use of machine learning methods have increased the complexity and scale of research pipelines, making code generation, refactoring, and debugging critical components of academic labor \citep{gentzkow2019text}. In this context, AI-assisted coding has emerged as one of the most widely adopted and potentially transformative applications of GenAI, enabling researchers to rapidly prototype, adapt, and scale empirical workflows. Evidence from controlled experiments in software development shows that access to coding assistants can materially reduce task completion time in realistic programming settings \citep{Peng2023_Copilot}. Beyond programming, field experimental evidence for knowledge workers documents sizable productivity and quality effects in tasks within the operational frontier of current models, while also emphasizing that performance can degrade when tasks fall outside that frontier, motivating structured human steering and workflow design \citep{BrynjolfssonLiRaymond2025_QJE, DellAcquaEtAl2023_JaggedFrontier}. Evidence from writing tasks similarly points to large productivity gains from access to LLMs in controlled settings \citep{NoyZhang2023_SSRN}.

Survey production provides a natural setting where these dynamics matter. Economic surveys are a core piece of the discipline's knowledge infrastructure, but their production is increasingly shaped by scale. The growth of working paper series and journal outlets has made it harder to sustain purely narrative synthesis without computational support. As a result, economists have progressively adopted text-as-data pipelines that provide a first-pass map of a research field and guide subsequent close reading \citep{ash2023textalgorithms,gentzkow2019text}. Within this toolkit, topic models, and in particular Latent Dirichlet Allocation (LDA), have become a pragmatic default for building quantitative taxonomies of large corpora. LDA offers an interpretable representation in which documents load on multiple themes and themes are defined by word distributions \citep{blei2003lda,blei2012probabilistic}. Applications in economics and adjacent areas use topic models precisely in a survey-like spirit: to describe the evolution of fields, to identify clusters, and to discipline what gets read in depth \citep{ambrosino2018topicmodeling,wehrheim2019economichistory}. The method has also become operationally accessible to economists through mainstream software environments \citep{schwarz2018ldagibbs}.

In parallel, a distinct body of work has developed around the environmental sustainability of artificial intelligence. The Green AI literature argues that progress in machine learning should be evaluated not only in terms of predictive performance, but also with respect to computational efficiency, energy consumption, and environmental impact \citep{schwartz2019greenai}. Foundational studies document the substantial carbon emissions associated with training and tuning large models, highlighting the role of extensive hyperparameter searches and repeated experimentation in driving computational costs \citep{strubell2019energy}. These contributions have been influential in establishing efficiency and transparency as central concerns in AI research.

Subsequent work has expanded the scope of Green AI beyond model training to include inference, deployment, and system-level considerations. Reviews and benchmarks show that, for large-scale and frequently used models, inference can account for a substantial share of total energy consumption, particularly when models are integrated into everyday workflows \citep{verdecchia2023greenai, jegham2025hungry}. This shift in focus has motivated the development of standardized methodologies and tools to measure and report the environmental footprint of computational tasks, including practical frameworks such as Green Algorithms and monitoring libraries such as \code{CodeCarbon} \citep{codecarbon_zenodo_2023, lacoste2019codecarbon}. 

Despite their shared relevance for contemporary research practice, the literatures on GenAI in economics and on Green AI have evolved largely in isolation. Studies on GenAI adoption focus primarily on productivity, research quality, and changes in academic labor, while abstracting from the environmental costs associated with AI-assisted workflows. Conversely, the Green AI literature concentrates on the efficiency of models and algorithms themselves, with limited attention to how researchers use AI systems as part of downstream scientific production processes. As a result, there is little systematic evidence on the environmental footprint of AI-assisted research activities in economics, particularly when GenAI acts as a coding and workflow tool rather than as the object of evaluation.

This paper contributes to the literature by explicitly linking these two strands at the level of research production. Rather than evaluating the energy efficiency of AI models in isolation, we study the computational footprint of an executed workflow produced with AI-assisted coding in a representative survey setting. By benchmarking alternative implementations of a large scale topic modelling pipeline under different prompting strategies, we provide reproducible evidence on how interaction design between researchers and generative systems affects runtime and estimated emissions. This motivates treating prompt design as a component of computational governance: a practical way to translate researcher choices about scope, stopping, and outputs into measurable differences in compute. The next section maps the recent Green AI literature to clarify what is well established and what remains fragmented, and we then introduce a controlled prompting benchmark that operationalises these mechanisms in a fixed environment and a decision-equivalent survey task.

\section{Mapping the Literature on Green AI}\label{sec:greenai_map}

Before turning to our benchmark of AI-assisted coding, we map different strands of the emerging Green AI literature to clarify which dimensions of efficiency and environmental impact are currently emphasised, and which receive comparatively less attention. This exercise serves two related purposes. First, it provides a systematic overview of a fast growing and heterogeneous body of work whose boundaries are often blurred with adjacent domains, such as AI applications for environmental or climate objectives. Second, it motivates our workflow focus by locating discussions of efficiency within the broader research landscape, and by highlighting the relative scarcity of evidence on downstream scientific workflows that increasingly rely on generative AI.

We construct a curated corpus of Green AI papers using the arXiv API combined with an explicit filtering protocol.\footnote{We used the following search terms: \texttt{"Green AI"}, \texttt{"carbon footprint" AND ("machine learning" OR "artificial intelligence")}, \texttt{"energy consumption of AI"}, \texttt{"energy efficiency" AND "deep learning"}, \texttt{"AI energy usage"}, \texttt{"environmental impact of large models"}, \texttt{"computational cost" AND "AI"}, \texttt{"efficient AI architectures"}, \texttt{"low-resource deep learning"}, \texttt{"sustainable AI"}.} Starting from 410 preprints retrieved through keyword and category based queries, we exclude papers that primarily apply AI to environmental or climate problems and retain those that explicitly study the environmental footprint, energy use, or efficiency of AI systems themselves. To operationalise this distinction at scale, we rely on a large language model classifier, in particular, ChatGPT 5.2, applied to titles and abstracts under consistent inclusion and exclusion rules (see Appendix~\ref{sec:lda_method} for the exact prompt and filtering protocol).\footnote{To rule out obvious misclassification patterns, we manually reviewed a random subset of 50 retained documents (titles and abstracts) to verify that the dominant contribution aligns with our Green AI definition, namely the environmental footprint or computational efficiency of AI systems, rather than AI applications to environmental problems.} The final dataset consists of 310 papers and is intended to capture methodological, algorithmic, and system oriented contributions rather than application focused work.\footnote{We focus on arXiv working papers because Green AI is a highly technical and rapidly evolving field, with a large share of contributions first disseminated in computer science venues. As documented in Appendix Figure~\ref{fig:corpus_evolution}, the number of Green AI relevant documents increases sharply in recent years, suggesting that restricting the corpus to peer reviewed outlets would substantially under represent current research activity. Moreover, arXiv is more representative than repositories such as SSRN for the computer science core of Green AI, where methodological and systems contributions are typically first circulated as preprints.}

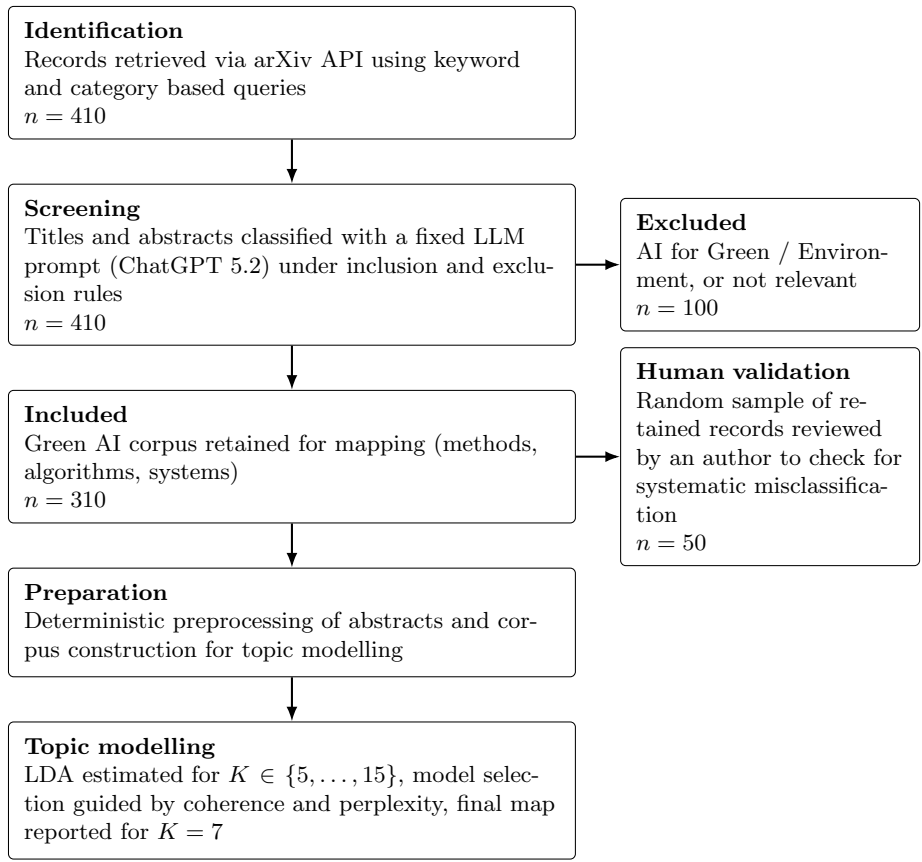
\begin{figure}[H]
\centering
\resizebox{\linewidth}{!}{%
\begin{tikzpicture}[
  font=\small,
  main/.style={rectangle, draw, rounded corners=2pt, align=left, inner sep=6pt, text width=0.60\linewidth},
  side/.style={rectangle, draw, rounded corners=2pt, align=left, inner sep=6pt, text width=0.30\linewidth},
  arrow/.style={-{Latex[length=2mm]}, thick},
  node distance=6mm
]

\node[main] (id) {\textbf{Identification}\\Records retrieved via arXiv API using keyword and category based queries\\$n=410$};

\node[main, below=of id] (screen) {\textbf{Screening}\\Titles and abstracts classified with a fixed LLM prompt (ChatGPT 5.2) under inclusion and exclusion rules\\$n=410$};

\node[side, right=6mm of screen] (excl) {\textbf{Excluded}\\AI for Green / Environment, or not relevant\\$n=100$};

\node[main, below=of screen] (incl) {\textbf{Included}\\Green AI corpus retained for mapping (methods, algorithms, systems)\\$n=310$};

\node[side, right=6mm of incl] (check) {\textbf{Human validation}\\Random sample of retained records reviewed by an author to check for systematic misclassification\\$n=50$};

\node[main, below=of incl] (prep) {\textbf{Preparation}\\Deterministic preprocessing of abstracts and corpus construction for topic modelling};

\node[main, below=of prep] (lda) {\textbf{Topic modelling}\\LDA estimated for $K\in\{5,\dots,15\}$, model selection guided by coherence and perplexity, final map reported for $K=7$};

\draw[arrow] (id) -- (screen);
\draw[arrow] (screen.east) -- (excl.west);
\draw[arrow] (screen) -- (incl);
\draw[arrow] (incl.east) -- (check.west);
\draw[arrow] (incl) -- (prep);
\draw[arrow] (prep) -- (lda);

\end{tikzpicture}%
}
\caption{Identification and screening pipeline for the Green AI literature map. Counts reflect our arXiv retrieval and filtering protocol.}
\label{fig:greenai_identification_pipeline}
\end{figure}

Once the corpus is set, we use topic models as a structured entry point into the Green AI literature, following a survey practice that has become increasingly common in economics when the volume of relevant material exceeds what can be covered through narrative synthesis alone, as it provides a transparent way of organising reading effort and documenting how broad research areas are delineated \citep{ash2023textalgorithms,ambrosino2018topicmodeling}. In particular, we apply Latent Dirichlet Allocation (LDA) \citep{blei2003lda} to identify latent themes in the Green AI abstracts. LDA is a generative probabilistic model in which documents are represented as mixtures of latent topics, and topics are represented as probability distributions over words. In this setting, LDA is used as a measurement device that summarises patterns of thematic co-occurrence in abstracts rather than as a tool to recover ``true'' topics in a strong ontological sense. Interpretation therefore relies on triangulation with close reading and domain knowledge.

Formally, each document $d$ is characterised by a topic proportion vector $\theta_d$, drawn from a Dirichlet prior, \(\theta_d \sim \text{Dirichlet}(\alpha)\), where the hyperparameter $\alpha$ governs the concentration of topics within documents. Each topic $k$ is in turn characterised by a word distribution $\phi_k$, \(\phi_k \sim \text{Dirichlet}(\beta)\), where $\beta$ controls the smoothness of topic--word distributions. For each token position $n$ in document $d$, a latent topic assignment $z_{d,n}$ is drawn from $\theta_d$, and the observed word $w_{d,n}$ is drawn from the corresponding topic distribution $\phi_{z_{d,n}}$. The estimated topic--word probabilities $p(w \mid k)$ provide the basis for topic labelling and interpretation (Appendix Table~\ref{tab:topic_tokens}).

Model selection is guided by two complementary diagnostics: topic coherence and perplexity. Topic coherence captures the semantic relatedness of the most probable words within a topic and is commonly used as a proxy for interpretability. A standard coherence measure can be written as
\begin{equation}
\text{Coherence}(k) = \sum_{i < j} \log \frac{D(w_i, w_j) + \epsilon}{D(w_i)} ,
\end{equation}
where $D(w_i, w_j)$ denotes the number of documents in which words $w_i$ and $w_j$ co-occur, and $\epsilon$ is a smoothing constant. Higher coherence values indicate more semantically consistent topics.

Perplexity, by contrast, is a likelihood-based measure that captures how well the model predicts held-out data. For a corpus of documents $\mathcal{D}$, perplexity is defined as
\begin{equation}
\text{Perplexity}(\mathcal{D}) = \exp\left(
-\frac{\sum_{d \in \mathcal{D}} \log p(w_d)}{\sum_{d \in \mathcal{D}} N_d}
\right),
\end{equation}
where $p(w_d)$ denotes the likelihood of the observed words in document $d$ under the model and $N_d$ is the number of tokens in $d$. Lower perplexity indicates better predictive fit. While perplexity is informative about statistical fit, it does not necessarily align with human interpretability, which motivates considering coherence and perplexity jointly when selecting the number of topics.

We now apply these diagnostics in our corpus to select a parsimonious topic representation for mapping the Green AI field. Abstracts are preprocessed using standard deterministic procedures, and we estimate topic models over a grid of topic counts. We focus on $K \in \{5,\dots,15\}$ because values below five yield topics that are too coarse to separate the main strands visible in close reading, while values above fifteen tend to fragment coherent themes into narrower subtopics with limited additional interpretive value in this dataset. This range therefore provides an informative window in which coherence and perplexity can reveal where marginal gains from increasing $K$ begin to diminish.

Within this grid, model selection combines quantitative diagnostics with qualitative interpretability. In practice, coherence improves markedly as $K$ increases up to seven topics and exhibits only limited gains thereafter, while higher-$K$ solutions become less semantically distinct upon inspection. Taken together, these considerations motivate the parsimonious $K=7$ specification used in the analysis, which balances interpretability and model fit.

Topic labels are assigned based on inspection of top terms and representative documents.\footnote{We also used the LDAvis relevance framework (via \code{pyLDAvis}) as an exploratory aid for topic interpretation; see \citet{sievert2014ldavis}. We do not report the interactive panels.} Because LDA is an unsupervised method, numeric topic identifiers are arbitrary. To ensure consistent presentation across tables and figures, we map internal topic IDs (0--6) to the presentation topics (1--7) by matching their token distributions (Appendix Table~\ref{tab:topic_id_mapping}). Detailed token distributions are reported in Appendix Table~\ref{tab:topic_tokens}, and compact visual summaries for each topic are provided in Appendix Figure~\ref{fig:wordclouds_topics}. Full methodological details are reported in Appendix~\ref{sec:lda_method}.

Figure~\ref{fig:coherence_perplexity} reports coherence and perplexity profiles across topic counts.

\begin{figure}[H]
    \centering
    \includegraphics[width=0.95\textwidth]{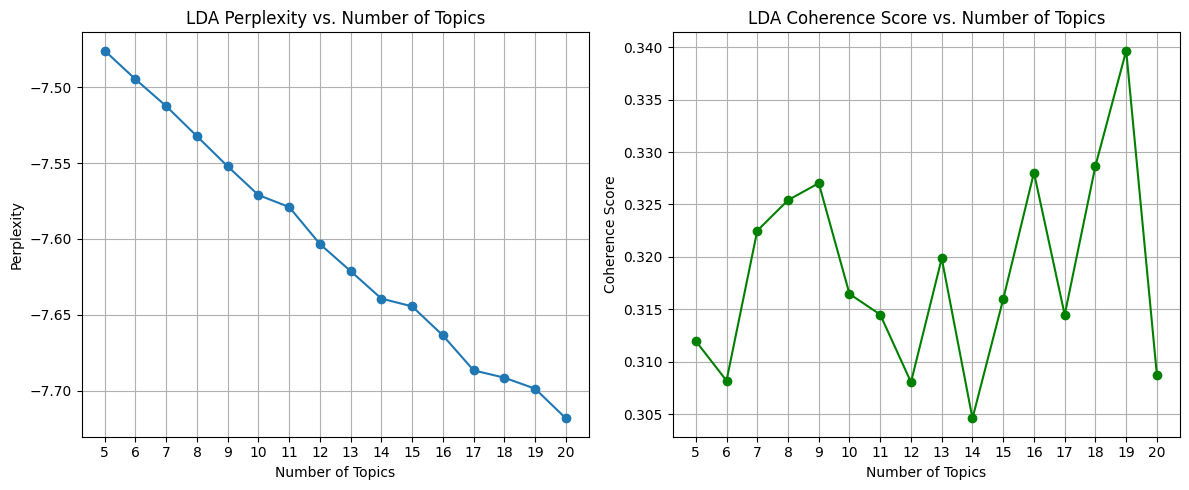}
    \caption{Model-selection diagnostics for the Green AI corpus: topic coherence ($c_v$, higher is better) and held-out perplexity (lower is better) for LDA models estimated over $K\in\{5,\ldots,15\}$. Each point corresponds to one fitted model on the preprocessed abstract corpus; the main specification uses $K=7$ based on the coherence plateau and interpretability checks.}
    \label{fig:coherence_perplexity}
\end{figure}

Table~\ref{tab:greenai_topics} summarises the seven topic clusters identified in our corpus and their relative prevalence. The most prominent topic concerns the carbon footprint and environmental impact of model training, accounting for just over one quarter of all tokens. Other sizeable clusters focus on inference and architectural optimisation, methods for measuring energy consumption, and system-level energy efficiency. Overall, these patterns point to a literature that concentrates heavily on computational efficiency at the level of models and infrastructure. By contrast, topics related to task specific green algorithms, broader sustainability narratives, and security--efficiency trade-offs appear less frequently, suggesting a more fragmented and specialised set of contributions.

\begin{table}[H]
\centering
\caption{Green AI topic clusters from the $K=7$ LDA model and their prevalence. “Share of tokens” reports the fraction of all preprocessed abstract tokens assigned to each topic under the fitted model (topics relabelled for presentation; see Appendix Table~\ref{tab:topic_id_mapping}).}
\label{tab:greenai_topics}
\begin{tabular}{@{}c >{\raggedright\arraybackslash}p{0.63\linewidth} c@{}}
\toprule
Topic & Label & Share of tokens (\%) \\
\midrule
1 & Carbon footprint and environmental impact of training & 28.1 \\
2 & System-level energy optimisation (hardware and infrastructure) & 12.1 \\
3 & Sustainability narratives, frameworks, and meta-research & 14.9 \\
4 & Energy consumption and environmental methods & 8.8 \\
5 & Inference and architecture optimisation for efficiency & 13.7 \\
6 & Green algorithms and task specific learning & 11.2 \\
7 & Carbon-aware AI and security efficiency trade-offs & 11.3 \\
\bottomrule
\end{tabular}
\end{table}

\begin{figure}[H]
    \centering
    \includegraphics[width=0.85\textwidth]{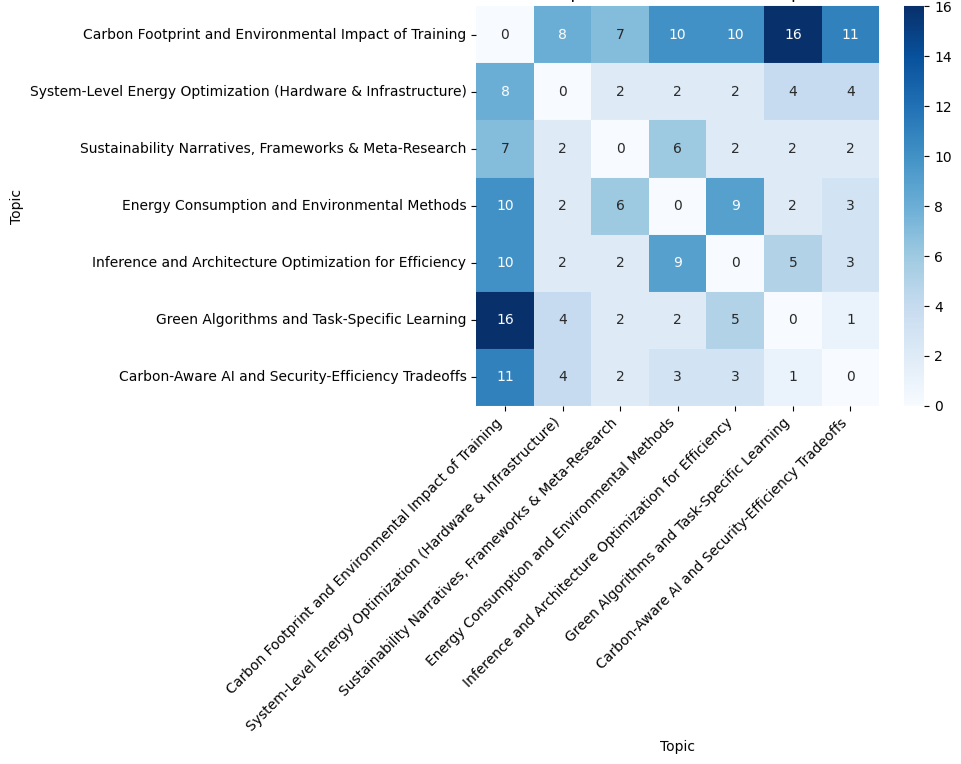}
    \caption{Topic co-occurrence in the Green AI corpus. Cells report the frequency with which two topics appear within the same document (based on document-level topic mixtures from the $K=7$ LDA model), highlighting which themes tend to co-appear in abstracts.}
    \label{fig:coocurrence}
\end{figure}

To characterise how these themes interact, we examine topic co-occurrence across documents. Figure~\ref{fig:coocurrence} reports the resulting heatmap, while Appendix Figure~\ref{fig:network} presents a network representation of the same structure.

\begin{figure}[H]
    \centering
    \includegraphics[width=0.95\textwidth]{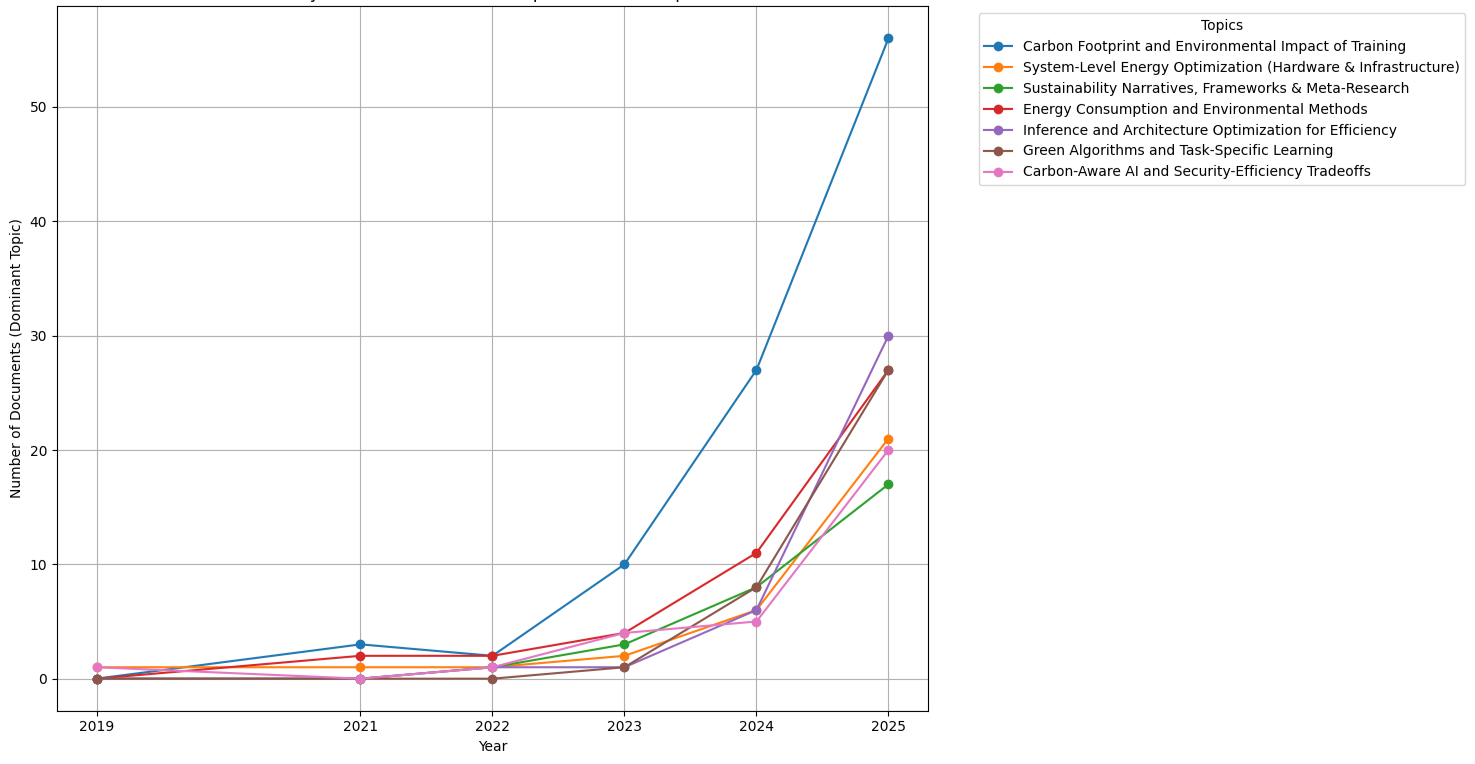}
    \caption{Yearly evolution of Green AI themes. For each year, we count documents by their dominant LDA topic (highest topic share within the document) using the $K=7$ model, showing how the composition of the retained arXiv corpus shifts over time.}
    \label{fig:topic_evolution}
\end{figure}

Figure~\ref{fig:topic_evolution} complements this static map by showing how the dominant topic composition evolves over time. Inference and architecture optimisation is among the fastest-growing strands, consistent with an increasing focus on deployment-stage constraints and efficiency concerns.

A first cluster centres on the footprint of training large models. Its vocabulary is dominated by terms such as \emph{training}, \emph{model}, \emph{data}, \emph{energy}, and \emph{carbon}, reflecting a research agenda concerned with the environmental costs of scale. Contributions in this area quantify the emissions implications of model development and experimentation, discuss how footprint estimates depend on hardware and location, and emphasise the need for transparent reporting to make results comparable across studies \citep{strubell2019energy,henderson2020reporting,schwartz2019greenai}.\footnote{All corresponding token distributions are reported in Appendix Table~\ref{tab:topic_tokens}. Compact word clouds for each topic are reported in Appendix Figure~\ref{fig:wordclouds_topics}.}

Closely related is a methodological cluster focused on how energy use and emissions are measured and reported. Rather than proposing efficiency improvements per se, this strand seeks to make carbon accounting operational through tracking tools, reproducible protocols, and practical reporting conventions \citep{lacoste2019codecarbon,codecarbon_zenodo_2023,henderson2020reporting}. Its proximity to the training-footprint cluster suggests that measurement infrastructure is increasingly treated as foundational, since conclusions about progress in ``greenness'' depend on whether footprint estimates are constructed consistently.

A third cluster shifts attention from models to infrastructure. References to \emph{hardware}, \emph{GPU}, and system performance point to work in which the primary margin of adjustment lies in the computing stack rather than in the learning algorithm. These studies analyse how energy use varies with hardware selection, utilisation rates, and system-level scheduling and provisioning, framing efficiency as an optimisation problem under operational constraints \citep{wang2025carbonaware,bian2023cafe,kadian2025tokenpowerbench}.

A fourth cluster focuses on inference and architectural efficiency. Here the emphasis is on reducing model size, compressing representations, or redesigning architectures to deliver comparable performance at lower computational cost. This strand is particularly relevant in applied settings where inference, rather than training, dominates lifecycle energy use once models are deployed at scale \citep{jegham2025hungry,li2024sustainablegenai,kadian2025tokenpowerbench}.

A fifth cluster groups what can be described as green algorithms in a narrower sense: changes to training and evaluation procedures that eliminate unnecessary computation. Typical examples include stopping rules, constrained experimentation, and task-specific simplifications that preserve the informational objective while reducing runtime \citep{schwartz2019greenai,zhong2024ssprop,ferdous2025green_prompting}. This cluster is conceptually closest to our empirical exercise, as it highlights how deliberate design choices and explicit decision rules can materially change computational intensity.

A sixth cluster is more explicitly conceptual and meta-scientific. It brings together work that discusses sustainability frameworks, evaluation criteria, and norms for assessing progress when efficiency is treated as a first-order objective rather than a side constraint \citep{schwartz2019greenai,verdecchia2023greenai,singh2024survey}. Rather than focusing on technical fixes, this literature addresses how incentives, benchmarks, and reporting practices shape research behaviour when accuracy remains the dominant performance metric.

Finally, a seventh cluster connects carbon awareness with broader system constraints, including security and operational considerations. The co-occurrence patterns suggest that efficiency is often discussed alongside limits on what can be shifted, simplified, or outsourced, so that emissions reductions emerge as part of a multi-objective decision problem rather than as a purely technical upgrade \citep{bian2023cafe,yousefpour2023greenfed,wang2025carbonaware}.

To make the thematic interpretation transparent and facilitate follow-up reading, Table~\ref{tab:topic_refs} reports a concise set of representative references for each Green AI topic cluster.

\begin{table}[H]
\centering
\caption{Green AI topic clusters: focus and representative references (illustrative, not exhaustive).}
\label{tab:topic_refs}
\renewcommand{\arraystretch}{1.15}
\begin{tabular}{c p{0.38\linewidth} p{0.48\linewidth}}
\toprule
Topic & Focus & Representative references \\
\midrule
1 & Carbon footprint and environmental impact of training &
\citep{strubell2019energy,henderson2020reporting,schwartz2019greenai}
\\[3pt]

2 & System-level energy optimisation (hardware and infrastructure) &
\citep{wang2025carbonaware,bian2023cafe,kadian2025tokenpowerbench}
\\[3pt]

3 & Sustainability narratives, frameworks, and meta-research &
\citep{schwartz2019greenai,verdecchia2023greenai,singh2024survey}
\\[3pt]

4 & Energy consumption and environmental methods (measurement and reporting) &
\citep{lacoste2019codecarbon,codecarbon_zenodo_2023,henderson2020reporting}
\\[3pt]

5 & Inference and architecture optimisation for efficiency &
\citep{jegham2025hungry,li2024sustainablegenai,kadian2025tokenpowerbench}
\\[3pt]

6 & Green algorithms and task specific learning &
\citep{schwartz2019greenai,zhong2024ssprop,ferdous2025green_prompting}
\\[3pt]

7 & Carbon-aware AI and security--efficiency trade-offs &
\citep{bian2023cafe,yousefpour2023greenfed,wang2025carbonaware}
\\
\bottomrule
\end{tabular}
\end{table}

Taken together, the topic structure points to a literature organised around two broad questions: how to measure the environmental footprint of AI systems in a credible and comparable way, and how to reduce that footprint through design choices that remain compatible with substantive research objectives \citep{schwartz2019greenai,verdecchia2023greenai}. At the same time, the mapping reveals a clear imbalance. While there has been substantial progress on measuring and reducing the footprint of model training, inference, and infrastructure, comparatively little attention has been paid to the environmental cost of downstream scientific workflows that are built and executed with generative AI assistance. The empirical analysis that follows addresses this gap by benchmarking the carbon footprint of AI-assisted coding in a representative research setting.

\section{A controlled prompting experiment}\label{sec:implementation}

This section provides a focused benchmark intended to complement the previous mapping of existing Green AI literature. The goal is not to estimate global footprint levels, but to illustrate a workflow-level mechanism and a replicable measurement template that economic survey authors can adapt.

\subsection{Prompts as decision policies}\label{sec:framing}

AI-assisted coding lowers the time and cognitive cost of producing and modifying code, which reduces the marginal cost of iteration and can increase research throughput. In a workflow setting, however, the environmental cost of this productivity gain depends on how the interaction is structured. When a researcher delegates implementation choices to a language model, the realised amount of computation is shaped by the prompt, because the prompt implicitly determines search scope, stopping behaviour, and which intermediate outputs are computed and stored.

We formalise this intuition by treating a prompting strategy as a decision policy. Let $p \in \mathcal{P}$ index alternative prompts. Let $Q(p)$ denote the practical value of the resulting workflow output, and let $C(p)$ denote its computational cost, measured by runtime or CO$_2$e. The researcher faces a simple trade off:

\begin{equation}
\label{eq:tradeoff_objective}
\max_{p \in \mathcal{P}} \; Q(p) - \lambda C(p),
\end{equation}

where $\lambda > 0$ captures the weight placed on computational cost. In this view, generic prompts can leave wide discretion to the system and increase redundant execution, raising $C(p)$ without improving $Q(p)$. By contrast, prompts that encode operational constraints and explicit decision rules can reduce $C(p)$ while preserving practically equivalent outputs.

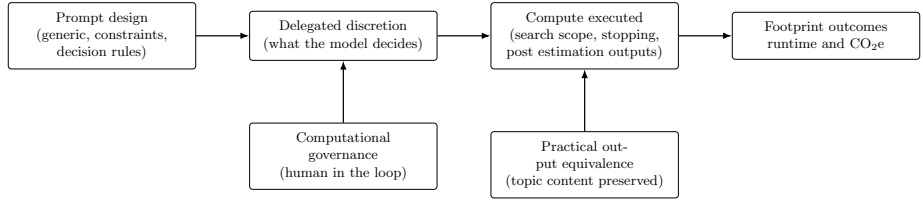
\begin{figure}[H]
\centering
\resizebox{\linewidth}{!}{%
\begin{tikzpicture}[
  font=\small,
  box/.style={rectangle, draw, rounded corners=2pt, align=center, inner sep=6pt, text width=38mm},
  sbox/.style={rectangle, draw, rounded corners=2pt, align=center, inner sep=6pt, text width=38mm},
  arrow/.style={-{Latex[length=2mm]}, thick}
]

\node[box] (prompt) {Prompt design\\(generic, constraints,\\decision rules)};
\node[box, right=12mm of prompt] (discretion) {Delegated discretion\\(what the model decides)};
\node[box, right=12mm of discretion] (compute) {Compute executed\\(search scope, stopping,\\post estimation outputs)};
\node[box, right=12mm of compute] (footprint) {Footprint outcomes\\runtime and CO$_2$e};

\draw[arrow] (prompt) -- (discretion);
\draw[arrow] (discretion) -- (compute);
\draw[arrow] (compute) -- (footprint);

\node[sbox, below=14mm of discretion] (gov) {Computational governance\\(human in the loop)};
\node[sbox, below=14mm of compute] (equiv) {Practical output equivalence\\(topic content preserved)};

\draw[arrow] (gov.north) -- (discretion.south);
\draw[arrow] (equiv.north) -- (compute.south);

\end{tikzpicture}%
}
\caption{Conceptual mechanism: prompt design allocates discretion between researcher and system, shaping search scope, stopping, and which outputs are computed, which determines executed compute and therefore runtime and CO$_2$e. Output equivalence is required to interpret footprint reductions as efficiency gains.}
\label{fig:mechanism}
\end{figure}

Figure~\ref{fig:mechanism} summarises this mechanism. Prompts affect the degree of delegated discretion, which in turn determines how much computation is executed through choices about search scope, stopping behaviour, and which post-estimation outputs are produced.

A key implication is that the cost channel is mediated by the amount of computation actually executed. We therefore interpret computational cost as

\begin{equation}
\label{eq:cost_decomposition}
C(p) = g\bigl(S(p), R(p), O(p), E\bigr),
\end{equation}

where $S(p)$ captures search scope, $R(p)$ captures stopping behaviour, $O(p)$ captures post estimation outputs, and $E$ denotes the (fixed) execution environment which is held constant in our experiment.\footnote{In Figure~\ref{fig:mechanism}, these components correspond to the links between delegated discretion and compute executed.}

The controlled design below operationalises this mechanism by generating the same LDA based workflow under alternative prompts and measuring how changes in delegated discretion translate into differences in runtime and CO$_2$e.

\subsection{Methodology and design}\label{sec:setup}

This subsection summarises the experimental architecture used to benchmark the computational footprint of AI-assisted coding under alternative prompting strategies. We define the treatments (prompting strategies), the outcome measures (runtime and CO$_2$e), the paired design used to reduce noise in a shared cloud environment, and the practical output equivalence test used to verify that footprint reductions do not reflect changes in the substantive object of analysis.

The experiment consists of generating an LDA-based literature-mapping pipeline with the assistance of a large language model, namely ChatGPT 5.2, under alternative prompting strategies and measuring the resulting algorithmic carbon footprint of the generated code. We begin with a generic prompt that requests a conventional, end-to-end implementation of an LDA pipeline and delegates most implementation and computational decisions to the language model. We then repeat the same task using three increasingly guided prompts that introduce different degrees of human steering. The first, which we label \emph{green soft}, adds high-level statements encouraging energy-efficient coding without imposing explicit constraints. The second introduces concrete \emph{computational restrictions} aimed at improving efficiency, while keeping the analytical objective unchanged. The third is \emph{decision-driven}, and reflects a more interventionist mode of human control in which the researcher specifies key decision rules and limits the scope of computation ex ante. The primary outcomes are total elapsed runtime (seconds) and estimated emissions (grams CO$_2$e) measured with \texttt{CodeCarbon}. Because the data, analytical objective, and execution environment are held fixed, observed differences in runtime and estimated emissions are plausibly attributable to the prompting strategy, up to residual noise from a shared cloud environment which we mitigate through blocked paired runs. This way, the experiment isolates the extent to which deliberate human steering in AI-assisted coding can reduce the environmental footprint of a decision-equivalent research task.\footnote{``Decision-equivalent'' means that all strategies are evaluated against the same analytical objective (selecting a plausible $K$ and producing the outputs needed to interpret the selected model), while differing in how exhaustively they explore intermediate specifications and which intermediate artefacts they compute and store.} Therefore, the contribution of the experiment is not methodological with respect to topic modelling itself, but empirical, building a setting which we treat as a representative example of the text-as-data workflows that increasingly underpin survey production in economics and finance.

The next subsection reports the main implementation choices that instantiate this design in a transparent and replicable workflow, while holding the task and environment fixed across prompting strategies.

\subsection{Implementation details}\label{sec:implementation_details}

This subsection describes the execution environment, data, and workflow implementation held constant across prompting strategies. These details support reproducibility and clarify which components of the footprint are included in the benchmark.

All pipelines are implemented in Python and executed in Google Colab under a fixed software environment.\footnote{We rely on the free Colab tier. As a shared computing environment, this can introduce run-to-run variation due to transient hardware allocation and background processes. To mitigate this source of noise, all comparisons are based on repeated paired runs executed back-to-back with counterbalanced order \citep{montgomery2017design}.} Library versions, random seeds, and execution settings are held constant across runs so that observed differences can be attributed to prompting choices rather than to changes in the software stack or stochastic variation in estimation.

Carbon emissions are tracked using \texttt{CodeCarbon}, an open-source monitoring library that estimates CO$_2$e emissions from computation by combining information on hardware utilisation with assumptions about energy consumption and the carbon intensity of electricity generation \citep{lacoste2019codecarbon, codecarbon_zenodo_2023}. In its default configuration, which we adopt here, \texttt{CodeCarbon} infers machine characteristics where possible and applies standard power and location-specific emissions factors when direct metering is unavailable. Because these estimates necessarily depend on assumptions about the energy mix and utilisation profiles, we complement emissions measurements with total elapsed runtime, which is recorded directly and provides a transparent and reproducible proxy for computational intensity.

Notably, our footprint accounting covers the computation executed by the survey workflow in the Colab runtime (preprocessing, model search/selection, and post-estimation outputs), measured by elapsed runtime and by \texttt{CodeCarbon}'s CO$_2$e estimates. This boundary excludes the data-centre footprint of LLM inference used for code generation and corpus filtering, which can be material but requires different instrumentation and is addressed in complementary work on LLM inference benchmarking \citep{jegham2025hungry,kadian2025tokenpowerbench,lannelongue2021green}.

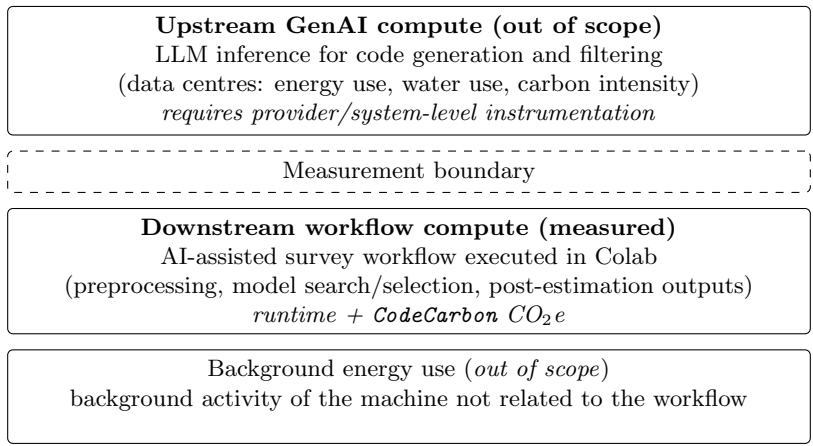
\begin{figure}[H]
\centering
\begin{tikzpicture}[
  font=\small,
  box/.style={draw, rounded corners=2pt, minimum width=0.88\linewidth, minimum height=0.9cm, align=center},
  sep/.style={draw, dashed, rounded corners=2pt, minimum width=0.88\linewidth, minimum height=0.55cm, align=center}
]
\node[box] (idle) {Background energy use (\textit{out of scope})\\
background activity of the machine not related to the workflow\\};
\node[box, above=2mm of idle] (workflow) {\textbf{Downstream workflow compute (measured)}\\AI-assisted survey workflow executed in Colab\\(preprocessing, model search/selection, post-estimation outputs)\\\textit{runtime + \texttt{CodeCarbon} CO$_2$e}};
\node[sep, above=2mm of workflow] (boundary) {Measurement boundary};
\node[box, above=2mm of boundary] (llm) {\textbf{Upstream GenAI compute (out of scope)}\\LLM inference for code generation and filtering\\(data centres: energy use, water use, carbon intensity)\\\textit{requires provider/system-level instrumentation}};
\end{tikzpicture}
\caption{Measurement boundary of the benchmark. We measure downstream workflow compute in Colab (runtime and \texttt{CodeCarbon} CO$_2$e) and exclude upstream LLM inference for code generation/filtering as well as background machine activity unrelated to the instrumented workflow.}
\label{fig:measurement_layers}
\end{figure}

Then, the empirical corpus used in our experiment consists of a large Scopus-derived dataset of research outputs in economics and finance that engage with artificial intelligence methods.\footnote{The dataset is constructed on 28 May 2025 by querying Scopus using a structured search that combines AI-related terms (including \textit{Generative AI}, \textit{Artificial Intelligence}, \textit{Machine Learning}, \textit{Natural Language Processing}, and \textit{Large Language Models}) with economic and financial terms (including \textit{Economics}, \textit{Finance}, \textit{Financial Planning}, and \textit{Investment}). The search is restricted to the subject areas of Decision Sciences, Business Management and Accounting, Economics, Econometrics and Finance, Social Sciences, and Energy. The query retrieves 19{,}185 records; after removing entries with missing, empty, or placeholder abstracts, the final corpus consists of 19{,}095 documents.} 
The unit of analysis is the abstract-level document, which provides a standardised and comparable representation of the studies included. Because computational cost in text-as-data pipelines scales with text length through tokenisation, preprocessing, and model estimation, we inspected abstract lengths after preprocessing. Abstracts are reasonably homogeneous in size, with most observations falling in a moderate range and only a small number of extreme cases. This regularity supports the use of abstracts as a consistent input for LDA and reduces the scope for runtime differences driven purely by variation in document length rather than by prompting strategy.

The analytical pipeline implemented in all experimental conditions is identical in substantive terms. Starting from the cleaned CSV file, the pipeline constructs document text by concatenating title and abstract when both are available, or using the abstract alone otherwise. Text preprocessing follows a deterministic sequence: lowercasing, removal of URLs, punctuation and digits, tokenisation, stopword removal, and lemmatisation using a fixed lemmatiser. Documents with fewer than five tokens after preprocessing are dropped. Topic models are then estimated using LDA with fixed hyperparameters and a fixed random seed across strategies, ensuring that any differences in computational outcomes cannot be attributed to stochastic variation or model tuning.

Beyond model estimation, the pipeline includes several post estimation steps that are standard in applied survey work but non trivial in computational terms. These include the computation of document--topic distributions, the construction of topic-level summary tables, and a time series representation tracking the prevalence of dominant topics over time. These components are explicitly included in the footprint under study, as they reflect how topic models are typically used in practice rather than in stylised benchmarking exercises.

\begin{figure}[H]
\centering
\begin{tikzpicture}[
  font=\small,
  node distance=6mm and 10mm,
  block/.style={rectangle, draw, rounded corners=2pt, align=center, inner sep=5pt, minimum width=42mm},
  io/.style={trapezium, trapezium left angle=70, trapezium right angle=110, draw, align=center, inner sep=5pt, minimum width=42mm},
  decision/.style={diamond, draw, align=center, inner sep=2pt, aspect=1.8},
  arrow/.style={-{Latex[length=2mm]}, thick},
  label/.style={font=\scriptsize, align=center}
]

\node[io] (data) {Input corpus\\CSV (title, abstract, year)};
\node[block, below=of data] (prep) {Preprocessing\\tokenise, stopwords, lemmatise};
\node[block, below=of prep] (train) {Fit LDA for candidate $K$\\(fixed hyperparams, seed)};
\node[block, below=of train] (score) {Compute coherence\\and log diagnostics};
\node[decision, below=of score] (stop) {Stop?};

\node[block, below=8mm of stop] (select) {Select $K^\ast$\\(rule depends on prompt)};
\node[block, below=of select] (post) {Post-estimation outputs\\topics, doc--topic, tables, trends};
\node[block, below=of post] (save) {Save outputs};

\draw[arrow]
  ([yshift=-1mm]stop.west) -- ++(-24mm,0)
  |- node[label, left, near start] {no} (train.west);

\draw[arrow] (data) -- (prep);
\draw[arrow] (prep) -- (train);
\draw[arrow] (train) -- (score);
\draw[arrow] (score) -- (stop);
\draw[arrow] (stop) -- node[label, right] {yes} (select);
\draw[arrow] (select) -- (post);
\draw[arrow] (post) -- (save);

\begin{pgfonlayer}{background}
\node[draw, dashed, rounded corners=3pt, fit=(prep)(train)(score)(stop)(select)(post)(save),
      inner sep=6pt, label={[label]north west:Measurement scope (runtime + \texttt{CodeCarbon})}] (meas) {};
\end{pgfonlayer}

\node[draw, rounded corners=2pt, align=left, font=\scriptsize,
      inner sep=3pt, right=6mm of meas.east] (tracknote)
{\textbf{Tracked:}\\
\texttt{CodeCarbon} $\rightarrow$ CO$_2$e\\
Runtime $\rightarrow$ elapsed time};

\end{tikzpicture}
\caption{Schematic workflow of the LDA pipeline and measurement scope. Candidate topic counts are evaluated by fitting LDA models and computing coherence diagnostics; the stopping rule and the choice of $K^\ast$ depend on the prompting strategy (e.g.\ exhaustive grid, coarse search, or decision-driven early stopping). Measurement starts at preprocessing and covers model search/selection and post-estimation computation (runtime and \texttt{CodeCarbon} CO$_2$e estimates).}
\label{fig:workflow}
\end{figure}
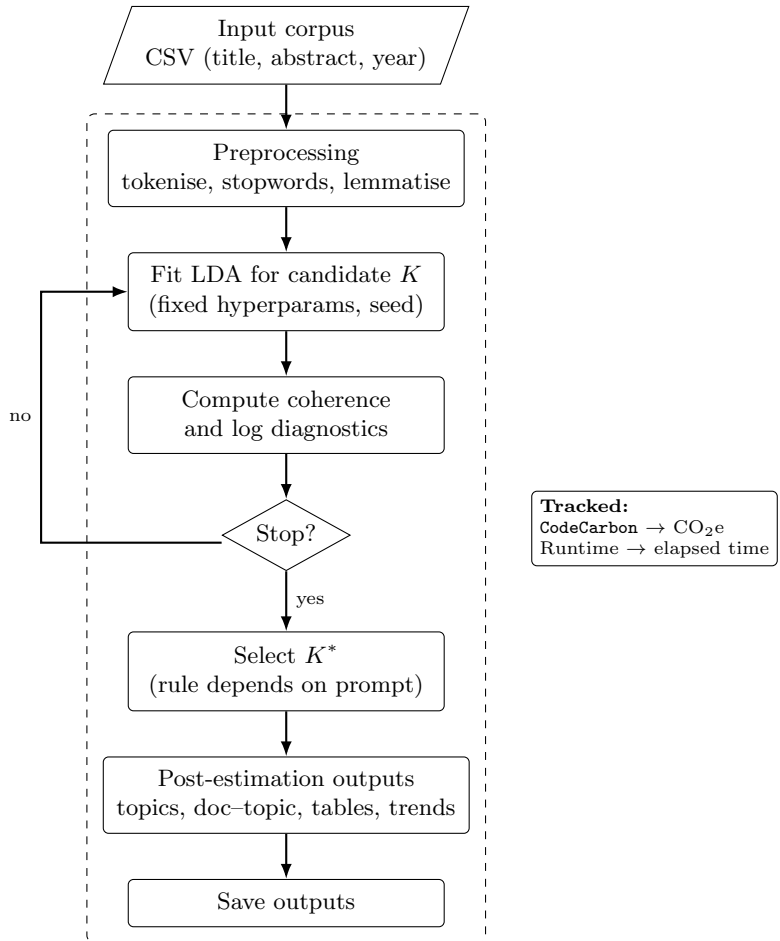

Figure~\ref{fig:workflow} provides a schematic overview of the workflow and highlights the stages at which computational cost and emissions are recorded.

We then set up a naive full prompt as the baseline condition of the experiment.\footnote{We interpret this baseline as an exhaustive, overproducing implementation that is realistic under naive prompting.} This prompt requests a complete, end-to-end implementation, while leaving a wide range of computational and organisational decisions implicit. In particular, choices related to the exploration of topic counts, the scope of post-estimation outputs, and the overall computational effort are not constrained ex ante, but are delegated to the language model. This interaction reflects a realistic mode of AI-assisted coding in which researchers rely on generative models to automate substantial parts of an empirical workflow without explicitly steering efficiency related trade-offs.

\begin{figure}[H]
\centering
\begin{tcolorbox}[
  colback=white,
  colframe=black,
  boxrule=0.4pt,
  width=0.9\textwidth,
  sharp corners,
  left=6pt,
  right=6pt,
  top=6pt,
  bottom=6pt
]
\scriptsize
\ttfamily
You are generating Python code to run in Google Colab.

\medskip
Assume the following variables already exist in the notebook:

\smallskip
CSV\_PATH (Path)\\
OUTPUT\_DIR (Path)

\medskip
Assume standard data science libraries are available
(pandas, numpy, gensim, nltk, matplotlib).

\medskip
Task:

\smallskip
Load the CSV from CSV\_PATH.\\
Use title + abstract as the text (if both are available; otherwise use abstract).\\
Preprocess the data.\\
Explore topic models by training LDA with different numbers of topics from 5 to 15 (inclusive).\\
For each model: print the topics.\\
For each model: compute and report a topic coherence score (c\_v).

\smallskip
For each trained model:\\
\quad Compute the document--topic distributions for all documents.\\
\quad Create a simple time series view showing how the number of documents per topic evolves over time.\\
\quad Produce a table that lists the most relevant documents for each topic.\\
Save the key outputs for each K to OUTPUT\_DIR so I can inspect them later
(including the coherence value).

\medskip
Use fixed random seeds where applicable (e.g., LDA random\_state=42)
so results are reproducible.

\medskip
The code should run end-to-end in one cell with no manual steps.\\
Return ONLY the Python code for that single cell.
\end{tcolorbox}
\caption{Generic prompt used to generate a baseline LDA pipeline with AI-assisted coding.}
\label{fig:generic_prompt}
\end{figure}

Building on this baseline, we introduce three alternative prompting strategies that explicitly encode different forms of human steering aimed at improving computational efficiency. These green prompts are defined and discussed below, with the full prompt templates reported in Appendix~\ref{app:prompts_green}.


The first strategy, labelled \emph{green soft prompting}, augments the baseline request with a generic instruction to write the code with an energy-efficient mindset. Importantly, this prompt does not impose explicit computational constraints, decision rules, or stopping criteria. Its purpose is to test whether high level appeals to efficiency, without operational guidance, are sufficient to induce meaningful changes in the generated code.

The second strategy, \emph{green computational prompting}, introduces explicit efficiency oriented constraints while preserving the same analytical objectives. E.g.: rather than evaluating every integer value of $K$ in the specified range, the prompt instructs the system to perform a coarse exploration of the topic grid within $[5,15]$ and to implement standard coding practices aimed at reducing redundant computation. The underlying model specification remains unchanged; only the breadth and organisation of the search over topic counts are affected.

The third strategy, \emph{ decision-driven stewardship}, formalises a trade-off that is familiar to applied researchers: balancing marginal analytical gains against computational cost. In this condition, the prompt instructs the system to evaluate topic models sequentially over a coarse grid and to apply an early stopping rule based on improvements in coherence. Once a preferred number of topics is selected, the pipeline computes computationally intensive post estimation outputs only for that selected model. This strategy explicitly encodes researcher judgement about what is worth computing, rather than treating all intermediate models symmetrically.

To mitigate noise arising from the shared Colab environment, a shared environment with time-varying hardware allocation and background load, we implement a paired experimental design. Each prompting strategy is evaluated in its own block of five paired repetitions against a freshly re-run naive baseline executed in the same session and counterbalanced order. The estimation is therefore the within-block paired difference (naive minus strategy), which mitigates transient noise but implies that baseline levels can differ across blocks. For this reason, relative savings (\%) are directly comparable across strategies, while absolute CO$_2$e savings in grams should be interpreted as conditional on the execution conditions and \texttt{CodeCarbon} assumptions within each block. In particular, for each green strategy, we run five paired repetitions against the naive baseline, executing the two pipelines back-to-back with counterbalanced order. The primary quantities of interest are the mean paired differences in total runtime and total emissions. This design allows us to attribute observed differences directly to prompting strategy rather than to transient system conditions.

Finally, because reductions in computational footprint are only substantively meaningful if the object of analysis remains unchanged, we explicitly test for output equivalence across prompting strategies. A direct comparison of topic models is non-trivial, as topic labels are not identified across runs. We therefore align topics across pipelines using an optimal matching procedure.

Topic models are compared at a common number of topics by first computing pairwise topic similarities and then solving an assignment problem that maximises total similarity across topics. Topic similarity is defined over the sets of top-$N$ words that characterise each topic. Let $W_i^{A}$ denote the set of top-$N$ words for topic $i$ under pipeline $A$, and $W_j^{B}$ the analogous set for topic $j$ under pipeline $B$. We measure similarity using the Jaccard index,

\begin{equation}
\label{eq:jaccard}
J_{ij} = \frac{|W_i^{A} \cap W_j^{B}|}{|W_i^{A} \cup W_j^{B}|},
\end{equation}

Stacking $J_{ij}$ into a $K \times K$ similarity matrix, we then solve the linear assignment problem

\begin{equation}
\label{eq:assignment}
\max_{\pi \in \mathcal{P}_K} \sum_{i=1}^{K} J_{i,\pi(i)},
\end{equation}

where $\mathcal{P}_K$ denotes the set of all permutations of $\{1,\dots,K\}$. This optimisation is implemented using the Hungarian algorithm.\footnote{The Hungarian algorithm \citep{munkres1957algorithms} is a classical method for solving linear assignment problems in polynomial time. It provides an optimal one-to-one matching between two sets given a pairwise similarity or cost matrix and is widely used to align unlabeled components across models, including topics in unsupervised text analysis.} The resulting matched similarities provide a transparent measure of whether different prompting strategies yield substantively equivalent topic content, allowing us to interpret differences in runtime and emissions as genuine efficiency gains rather than changes in the underlying clustering of documents.

Table~\ref{tab:external_validity} summarises what is held constant in the experiment, what may change in applied settings, and how such variation is expected to affect footprint levels and interpretation.

\begin{table}[H]
\centering
\caption{Experimental controls and external validity considerations.}
\label{tab:external_validity}
\begin{tabular}{@{}p{0.30\linewidth} p{0.34\linewidth} p{0.34\linewidth}@{}}
\toprule
Held constant in the experiment & May change in applied settings & Expected implication for footprint and interpretation \\
\midrule
Task and workflow structure (LDA based survey pipeline) & Other workflows (regressions, simulations, machine learning, data cleaning, writing with code execution, agentic pipelines) & Absolute levels will differ, but the mechanism should strengthen when workflows involve wider search spaces and more optional outputs, since discretion has more scope to expand compute. \\
Execution environment (Google Colab free tier) & Local machines, dedicated servers, cloud instances with different hardware and background load & Baseline runtime and emissions may shift, and noise may change, but decision rules should still reduce redundant execution when compute is unconstrained. \\
Model used for coding (ChatGPT 5.2) & Other models and tools (different LLMs, code copilots, multi agent tools) & More capable or more agentic systems can increase variance in compute when unconstrained. Governance via explicit rules should remain relevant as capability rises. \\
Measurement approach (runtime and CodeCarbon CO$_2$e) & Alternative carbon accounting assumptions, metered energy, different carbon intensity estimates & CO$_2$e levels can vary with assumptions, which motivates reporting runtime alongside emissions. Relative differences across prompting strategies should be more stable than levels. \\
\bottomrule
\end{tabular}
\end{table}

This clarifies the scope of the design. Our goal is not to claim universal footprint levels, but to identify a mechanism that is relevant across AI-assisted research workflows: prompt design structures delegated discretion, which in turn shapes how much computation is executed. By holding the task, model, and environment fixed, we can attribute differences in runtime and CO$_2$e to prompting strategies rather than to changes in data or infrastructure. 

Before presenting the empirical results, we interpret the prompting treatments through the decision-policy framework and derive testable implications for how each strategy should affect runtime and CO$_2$e.

\subsection{Interpretation and testable implications}\label{sec:interpretation}

Our framing in Section~\ref{sec:framing} treats prompting strategies as decision policies that trade off practical analytical value against computational cost. A more laissez-faire (naive) prompt typically increases the scope of model exploration and the number of intermediate artefacts computed and stored. This can raise the practical value of the workflow output, $Q(p)$, by generating additional diagnostics, richer documentation, and more candidate specifications to inspect. At the same time, broader exploration and additional outputs mechanically increase computational cost, $C(p)$, through longer runtimes and higher CO$_2$e. By contrast, prompts that encode operational constraints and explicit decision rules restrict delegated discretion and can reduce $C(p)$ without necessarily lowering $Q(p)$ in a practically meaningful way, if the removed computation is largely redundant or weakly informative.

Formally, this intuition corresponds to the objective in Eq.~\eqref{eq:tradeoff_objective} and the cost decomposition in Eq.~\eqref{eq:cost_decomposition}. Naive prompting tends to increase search scope $S(p)$ (e.g., evaluating a dense grid of topic counts and computing outputs for each candidate model), may delay stopping behaviour $R(p)$ (e.g., continuing evaluation even when improvements are small), and often expands post-estimation outputs $O(p)$ (e.g., generating full document--topic matrices and tables for every candidate $K$). Constraint-based prompting reduces one or more of these margins by design. In our experiment, this implies that differences in runtime and emissions should be interpretable as differences in the realised amount of computation executed, rather than as differences in the underlying analytical task.

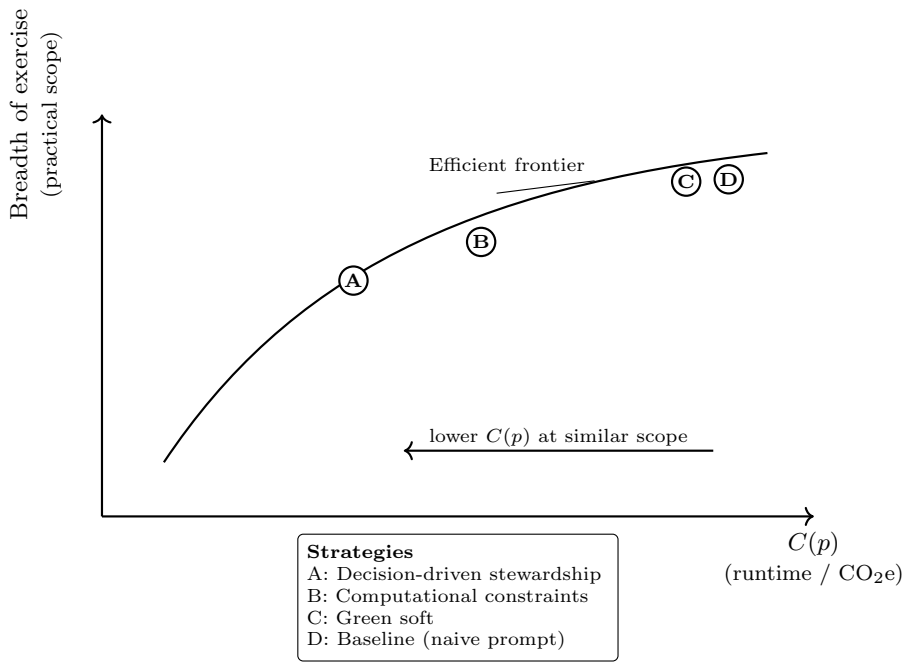
\begin{figure}[H]
\centering
\resizebox{\linewidth}{!}{%
\begin{tikzpicture}[font=\small, x=1cm, y=1cm]

  \draw[->, thick] (0,0) -- (9.2,0);
  \draw[->, thick] (0,0) -- (0,5.2);

  \node[align=center] at (9.2,-0.55) {$C(p)$\\{\footnotesize (runtime / CO$_2$e)}};
  \node[rotate=90, align=center] at (-0.85,5.2) {Breadth of exercise\\{\footnotesize (practical scope)}};

  \draw[thick]
    (0.8,0.7)
    .. controls (2.4,3.1) and (4.7,4.25) .. (8.6,4.70);

  \coordinate (frontpt) at (5.10,4.18);
  \node[anchor=west, font=\scriptsize] (frontlab) at (4.10,4.55) {Efficient frontier};
  \draw[-, thin] (frontlab.south east) -- (frontpt);

  \coordinate (A) at (3.25,3.05);  
  \coordinate (B) at (4.90,3.55);  
  \coordinate (C) at (7.55,4.33);  
  \coordinate (D) at (8.10,4.36);  

  \foreach \p/\lab in {A/A,B/B,C/C,D/D} {
    \draw[thick, fill=white] (\p) circle (5.2pt);
    \node[font=\bfseries\scriptsize] at (\p) {\lab};
  }

  \draw[->, thick] (7.9,0.85) -- (3.9,0.85);
  \node[font=\scriptsize, align=center] at (5.9,1.02)
    {lower $C(p)$ at similar scope};

  \node[draw, rounded corners=2pt, fill=white, font=\scriptsize, align=left]
    at (4.6,-1.05) {%
      \begin{tabular}{@{}l@{}}
      \textbf{Strategies}\\
      A: Decision-driven stewardship\\
      B: Computational constraints\\
      C: Green soft\\
      D: Baseline (naive prompt)\\
      \end{tabular}
    };

\end{tikzpicture}%
}
\caption{Stylised efficiency frontier for prompting strategies. The efficient frontier represents the maximum attainable breadth of exercise (practical scope) for a given computational cost $C(p)$ (runtime or CO$_2$e). Weak guidance (green soft) may leave outcomes below the frontier with little change relative to the naive baseline. Constraint-based prompting reduces cost but can also reduce breadth of exercise when the prompt narrows what is computed. Decision-driven prompting can further lower cost by restricting computation to decision-relevant outputs. Points are illustrative and not estimated.}
\label{fig:efficient_frontier}
\end{figure}

Figure~\ref{fig:efficient_frontier} summarises this interpretation using a stylised efficiency frontier. The vertical axis represents an operational proxy for practical value, namely the breadth of exercise (practical scope), which we denote $Q(p)$, while the horizontal axis represents computational cost $C(p)$ measured by runtime or CO$_2$e. The curve depicts the set of (conceptual) Pareto-efficient prompting strategies that achieve the highest attainable $Q$ for a given cost. A naive, unconstrained prompt may lie to the right of the frontier if it executes redundant computation that adds little analytical value, while prompts that encode constraints and decision rules can move the workflow leftwards (lower $C$) with little or no reduction in $Q$ when outputs remain practically equivalent.

This interpretation yields testable implications that guide the empirical analysis. First, generic ``green'' language without operational guidance should have limited effect if it does not systematically change $S(p)$, $R(p)$, or $O(p)$, implying small or unstable differences in runtime and emissions relative to the naive baseline. Second, computational-constraint prompting should deliver sizeable and stable footprint reductions by directly narrowing $S(p)$ and reducing redundant model estimation. Third, decision-driven prompting should yield the largest reductions by combining a restricted search procedure with an explicit stopping rule and by limiting expensive post-estimation computation to the selected specification, directly reducing both $R(p)$ and $O(p)$. Finally, because different prompting strategies remove computation in different phases of the workflow, reductions in runtime and reductions in CO$_2$e need not be proportional, motivating the joint reporting of both metrics in the results section.

\subsection{Results}\label{sec:results}

\begin{figure}[H]
\centering
\includegraphics[width=\linewidth]{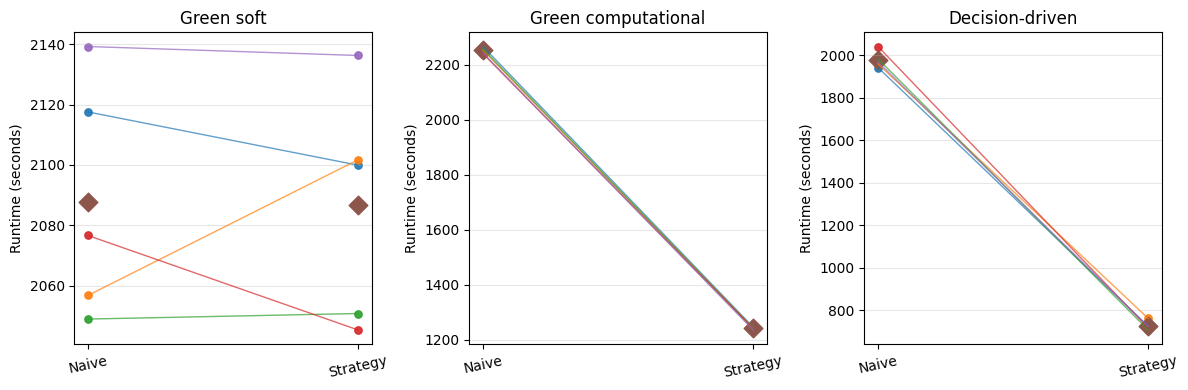}
\caption{Absolute runtime (seconds) by paired repetition. Each panel shows a blocked comparison between a freshly re-run naive baseline (Naive) and the corresponding prompting strategy (Strategy) for $n=5$ paired runs; lines connect within-pair outcomes.}
\label{fig:runtime_pairs}
\end{figure}

\begin{figure}[H]
\centering
\includegraphics[width=\linewidth]{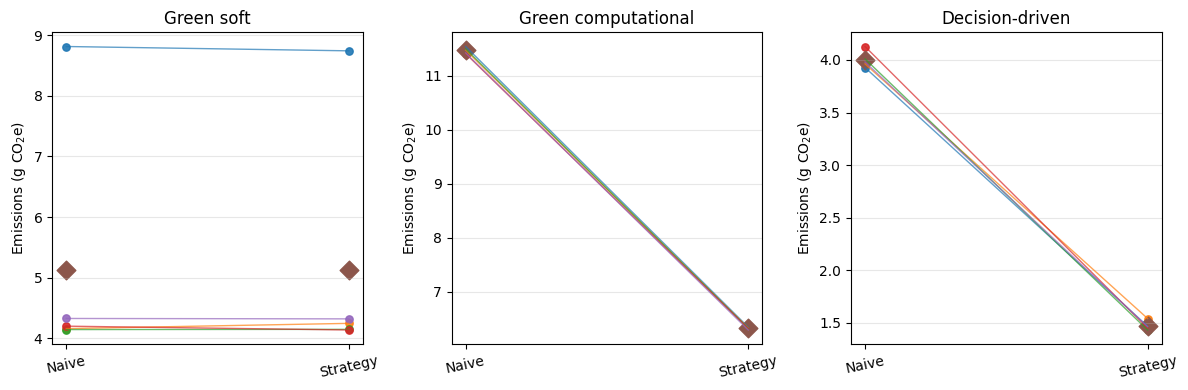}
\caption{Absolute emissions (g CO$_2$e) by paired repetition. Each panel shows a blocked comparison between a freshly re-run naive baseline (Naive) and the corresponding prompting strategy (Strategy) for $n=5$ paired runs; lines connect within-pair outcomes.}
\label{fig:emissions_pairs}
\end{figure}

We begin by comparing runtime and estimated carbon emissions across prompting strategies. Table~\ref{tab:results_main} reports mean paired savings in runtime and estimated emissions based on $n=5$ paired repetitions per strategy. To visualise the within-block comparisons, Figures~\ref{fig:runtime_pairs} and \ref{fig:emissions_pairs} plot absolute outcomes for each paired repetition, connecting each strategy run to its freshly re-run naive baseline executed in the same Colab session.

All comparisons are computed within blocked paired runs: each prompting strategy is evaluated against a naive baseline re-run in the same session and in counterbalanced order. This design mitigates time-varying noise in a shared cloud environment and isolates the computational effect of prompt design through within-block paired differences. Statistical inference is based on paired $t$-tests of within-block differences (Appendix~\ref{app:ttests}). Given the small number of paired repetitions ($n=5$), we complement these tests with non-parametric bootstrap confidence intervals for mean and median paired savings, and with an exact sign test based on the direction of paired savings (Appendix~\ref{app:robustness_paired}). 

\begin{table}[H]
\centering
\caption{Average paired savings in runtime and estimated emissions relative to a within-block naive baseline ($n=5$ paired runs per strategy).}
\label{tab:results_main}
\begin{tabular}{lcc}
\toprule
\multicolumn{3}{l}{\textbf{Runtime}} \\
\midrule
Strategy & Absolute savings (s) & Relative (\%) \\
\midrule
Green soft & 1.03 & 0.05 \\
Green computational constraints & 1011.07 & 44.89 \\
Decision-driven stewardship & 1251.39 & 63.26 \\
\midrule
\multicolumn{3}{l}{\textbf{Carbon emissions}} \\
\midrule
Strategy & Absolute savings (g CO$_2$e) & Relative (\%) \\
\midrule
Green soft & 0.010 & 0.20 \\
Green computational constraints & 5.149 & 44.90 \\
Decision-driven stewardship & 2.532 & 63.27 \\
\bottomrule
\end{tabular}
\begin{flushleft}\footnotesize \textit{Notes:} Each strategy is evaluated in a blocked paired design against a freshly re-run naive baseline in the same Colab session (counterbalanced order). Relative savings (\%) are comparable across strategies; absolute CO$_2$e savings (g) depend on \texttt{CodeCarbon} assumptions and transient execution conditions within each block. Positive values indicate lower runtime or emissions than the within-block baseline.\end{flushleft}
\end{table}

A first result is that the green soft prompt does not produce systematic reductions in computational cost.\footnote{Paired t-tests relative to the naive baseline fail to reject the null of zero mean difference for both runtime and carbon emissions (see Appendix~\ref{app:ttests}).} Figures~\ref{fig:runtime_pairs} and \ref{fig:emissions_pairs} show non-negligible dispersion in absolute outcomes across repetitions in this shared cloud environment, while the within-block paired differences (naive minus strategy) isolate the effect of prompt design. In other words, adding generic statements encouraging energy efficiency, without translating them into explicit operational guidance, does not reliably alter how the language model structures the analysis.\footnote{Occasional high-emission executions are visible in the run-level outcomes, but they do not translate into systematic paired savings relative to baseline.} This finding suggests that environmentally preferable outcomes do not arise automatically from vague normative instructions alone. Consistent with this, the paired $t$-tests fail to reject zero mean savings for both runtime and emissions (Appendix~\ref{app:ttests}).

However, introducing explicit computational constraints, while keeping the analytical objective unchanged, leads to large and highly stable efficiency gains. Relative to the naive baseline, the green computational constraints prompt reduces average runtime by approximately 1,000 seconds, corresponding to a reduction of about 45\%, and achieves a comparable reduction in carbon emissions. As shown in Table~\ref{tab:results_main}, these paired differences are remarkably consistent across runs, with very low dispersion. Figures~\ref{fig:runtime_pairs} and \ref{fig:emissions_pairs} show that the within-block paired differences are tightly clustered across repetitions, consistent with the very small standard deviations reported in Appendix~\ref{app:ttests} and with the non-parametric robustness checks reported in Appendix~\ref{app:robustness_paired}. This indicates that constraining the search space and eliminating redundant model estimation can substantially reduce computational cost, even when the overall structure of the pipeline and the scope of post estimation outputs remain unchanged. Paired $t$-tests strongly reject the null of zero mean savings for both runtime and emissions ($p<0.001$ in both cases; Appendix~\ref{app:ttests}).

While runtime and estimated emissions move in the same direction within each paired run, the conversion from seconds to grams of CO$_2$e is not invariant in a shared cloud environment. Different prompting strategies reduce computation in different phases of the workflow, and these phases can differ in average power draw (e.g., sustained model estimation versus lighter post-processing and I/O). In addition, each strategy is evaluated in a separate blocked paired experiment with a freshly re-run baseline, so baseline emissions per second can differ across blocks due to transient Colab conditions (hardware allocation, background load) and \texttt{CodeCarbon}'s inferred power draw and carbon-intensity assumptions. For this reason, we interpret cross-strategy patterns primarily through relative (\%) savings and paired differences, while treating absolute savings in grams as conditional on execution conditions within each block. This also helps explain why decision-driven prompting yields the largest time savings yet smaller absolute CO$_2$e savings than computational constraints in our table: the relevant baseline block exhibits a lower implied emissions intensity (grams per second).

The largest and most consistent reductions are observed under the decision driven prompting strategy. By encoding explicit stopping rules and restricting computationally intensive post estimation steps to the selected model, this approach delivers substantial savings relative to a freshly re run naive baseline executed within the same session. Figure~\ref{fig:runtime_pairs} shows that, across the five paired repetitions, runtimes drop sharply and uniformly when moving from the naive baseline to the decision driven workflow. Figure~\ref{fig:emissions_pairs} displays the same pattern for estimated emissions. These reductions are not driven by isolated executions: the paired lines show that, within each repetition, the strategy shifts to a systematically lower level than its within block baseline.

Computational constraints also reduce footprint markedly. Figures~\ref{fig:runtime_pairs} and~\ref{fig:emissions_pairs} show a clear downward shift from the naive baseline to the strategy in every paired repetition under computational constraints, indicating that limiting exploration and avoiding recomputation translates directly into lower runtime and emissions. By contrast, the green soft strategy exhibits little systematic change relative to its within-block baseline, consistent with near-zero paired savings.

Statistical inference is conducted at the level of within block paired differences, \(\Delta_i = Y^{\text{naive}}_i - Y^{\text{strategy}}_i\), for \(n=5\) paired repetitions. Paired \(t\) tests strongly reject the null of zero mean savings for decision driven stewardship and for green computational constraints in both runtime and emissions (\(p<0.001\)). For green soft, the null cannot be rejected. Full test statistics are reported in Appendix~\ref{app:ttests}. Given the small number of paired repetitions, Appendix~\ref{app:robustness_paired} reports bootstrap confidence intervals and an exact sign test as complementary, assumption-light diagnostics.

Importantly, the decision driven strategy does not rely on additional algorithmic sophistication or more advanced modelling techniques. Instead, it formalises a decision logic that many researchers apply informally when working interactively: deciding when further exploration yields diminishing returns and which outputs are worth computing. Making this logic explicit at the prompting stage therefore has first-order implications for the computational and environmental footprint of AI-assisted workflows.

A natural concern is that reductions in runtime and emissions might reflect changes in the analytical object rather than genuine efficiency gains. We therefore verify topic-output consistency across prompting strategies at a common specification, $K=7$, which is selected by the decision-driven strategy and provides a shared benchmark. Topics are compared using the top 10 words per topic; because topic labels are not identified, we compute pairwise similarities and apply the Hungarian algorithm to obtain the one-to-one assignment that maximises total similarity. Similarity is measured by the Jaccard index over sets of top words. In our controlled design, preprocessing, hyperparameters, and the random seed are held fixed,\footnote{Topic models are not identified in general, and different random initialisations can yield alternative local optima and topic labelings. We fix the random seed to isolate the effect of prompting strategies on computational footprint, rather than conflating it with stochastic estimation variation.} so $\overline{J}=1$ is expected and is interpreted as a sanity check that the analytical pipeline is unchanged across prompting strategies. Table~\ref{tab:topic_similarity} reports the resulting matched similarities. Consistently, the exported document--topic matrices and the top documents per topic match at $K=7$ in our logged outputs.

\begin{table}[H]
\centering
\caption{Sanity check of topic-output consistency relative to the naive baseline. Topics are compared at $K=7$ using the top 10 words per topic. Because topic labels are not identified across runs, topics are aligned using the Hungarian algorithm to maximise total similarity across topics. Similarity is measured by the Jaccard index over sets of top words. Under fixed seeds and identical specifications, $\overline{J}=1$ is expected and is interpreted as a verification that the analytical pipeline is unchanged across prompting strategies.}
\label{tab:topic_similarity}
\begin{tabular}{lc}
\toprule
Comparison (vs. naive) & Average matched Jaccard $\overline{J}$ \\
\midrule
Green soft & 1.000 \\
Green computational constraints & 1.000 \\
Decision-driven (chosen $K^\ast=7$) & 1.000 \\
\bottomrule
\end{tabular}
\end{table}

Taken together, the results support a human-in-the-loop interpretation of green AI practices. Generic exhortations to “be efficient” are insufficient to alter computational outcomes. In contrast, prompting strategies that impose explicit computational constraints already deliver substantial efficiency gains, and these gains become even larger when prompts encode decision rules that reflect researcher judgement about what is worth computing. The environmental footprint of AI-assisted research thus depends not only on model choice or hardware, but on how researchers articulate and enforce their analytical priorities when delegating tasks to generative systems.

The following section discusses the implications of these findings for research practice, reporting standards, and the broader Green AI agenda in economics.

\section{Discussion and implications}\label{sec:conclusions}

This paper addresses a simple research question with broad implications for economic research practice: when economists use GenAI to write code, do the resulting productivity gains come with an avoidable environmental cost, and can human steering mitigate that cost without altering the substantive output? Our controlled experiment delivers a clear message. Generic, high-level appeals to ``be efficient'' do not reliably change computational outcomes. By contrast, prompts that encode concrete constraints and, in particular, explicit decision rules generate large and robust reductions in runtime and CO$_2$e, while preserving identical topic outputs under optimal matching. In this setting, the environmental footprint of AI-assisted coding is therefore not an inherent property of using LLMs, but a consequence of how analytical discretion is allocated between the researcher and the model.

To situate these findings, it helps to place them alongside the emerging evidence on GenAI and scientific productivity. A growing body of work shows that GenAI can substantially raise output, including in economics and closely related social sciences. Recent work by \citet{filimonovic2025can} provides particularly compelling evidence in this regard, documenting sizable increases in publication output among GenAI adopters, alongside modest but positive effects on journal quality. These gains appear especially pronounced for early-career researchers, authors in technically demanding fields, and scholars based in non-English speaking countries. In parallel, large-scale evidence in the natural sciences suggests that individual gains can coexist with a collective narrowing of scientific attention, consistent with a contraction in topical breadth and lower follow-on engagement \citep{hao2025_ai_tools_nature}. Taken together, this evidence implies that generative AI lowers key frictions in scientific production, notably those related to writing, coding, and navigating technical complexity, thereby increasing throughput and researcher satisfaction, but these gains may also reshape what gets studied by changing the marginal cost of producing and iterating on ideas.

For economists, the productivity channel is particularly salient in survey production and text-as-data workflows. Scale increasingly constrains these projects: the bottleneck is no longer only reading and interpretation, but also the construction, iteration, and maintenance of computational pipelines that discipline what gets read. LLM-based coding assistance can relax this constraint by reducing the time and cognitive cost of writing and modifying code. In that sense, our findings align with the productivity narrative: AI-assisted coding works, it delivers usable pipelines, and it enables researchers to do more.

At the same time, our results qualify this optimistic view along a dimension that the economics literature has discussed less directly: computational efficiency. When the baseline prompt delegates too many decisions to the model (for example, how exhaustively to search over specifications, which intermediate outputs to compute, and what to store), the resulting pipeline can expend substantial computation that a human researcher would typically avoid. Productivity gains remain real, but they can come with unnecessary increases in runtime and emissions unless decision logic becomes explicit. In practical terms, generative AI reduces the marginal cost of iteration, which encourages experimentation, yet it can also amplify redundant execution in the absence of clear constraints. Table~\ref{tab:productivity_efficiency} summarises this productivity--efficiency trade-off for GenAI-assisted survey workflows and highlights the distinct margins on which human judgement governs both research output and computational footprint.

\begin{table}[H]
\centering
\caption{GenAI-assisted survey workflows: productivity gains and computational-efficiency risks.}
\label{tab:productivity_efficiency}
\begin{tabular}{@{}p{0.24\linewidth} p{0.36\linewidth} p{0.36\linewidth}@{}}
\toprule
 & \textbf{Productivity (research output)} & \textbf{Efficiency (computational footprint)} \\
\midrule
Primary margin
& Lower time and cognitive costs for producing, refactoring, and debugging code; faster iteration
& Runtime, energy use, and CO$_2$e of executed workflows, including optional search and post-estimation outputs \\
\midrule
What GenAI changes
& Reduces the marginal cost of iteration and enables more rapid prototyping of survey pipelines
& Can widen the realised search scope and expand computed artefacts when discretion is not constrained \\
\midrule
Main opportunity
& Faster construction and updating of computational surveys under scale
& Scope to reduce avoidable computation through explicit decision rules and stopping criteria \\
\midrule
Main risk (if unmanaged)
& ``More iterations'' may shift effort toward exploration and complexity without proportional analytical gains
& ``More automation'' may translate into redundant execution, inflated runtime, and higher emissions \\
\midrule
What our benchmark suggests
& Prompting strategies preserve the ability to generate end-to-end pipelines
& Generic green language has limited and unstable effects; operational constraints and decision-driven prompts can materially reduce runtime and estimated CO$_2$e in this workflow \\
\midrule
Role of human judgement
& Shapes research direction, interpretation, and what is read in depth
& Encodes search scope, stopping behaviour, and which outputs are decision-relevant (computational governance) \\
\bottomrule
\end{tabular}
\end{table}

From this perspective, the largest efficiency gains in our experiment arise not from new algorithms or more sophisticated modelling techniques, but from formalising a familiar research practice: stopping the search when marginal analytical gains become small, and restricting computationally intensive post-estimation outputs to the selected specification. This finding resonates with broader discussions about how LLMs may reshape research norms as production scales. As \citet{kusumegi2025scientific} argue, traditional signals of research quality, such as surface-level complexity, may become less informative in an era of AI-assisted writing and coding. Our results suggest a parallel implication on the computational side. Prompt design should function as a component of research governance, akin to coding standards or pre-analysis plans, because it determines how much discretion the model exercises and how much computation is deemed worth doing.

Related perspectives in the experimental methods literature make a similar point in a different register. LLMs can accelerate design and implementation, but researchers must retain the ability to interrogate assumptions, understand trade-offs, and document decisions \citep{charness2025next}. Our contribution provides a concrete computational analogue. Efficiency gains do not arise automatically from access to an LLM. They require researcher judgement translated into explicit constraints. In this sense, human-in-the-loop is not merely a normative slogan, but a measurable determinant of runtime and emissions in applied workflows.

A further implication concerns measurement and reporting. Although runtime and estimated CO$_2$e reductions move in the same direction in our paired runs, the conversion from time to emissions is not fixed in a shared cloud environment. Estimated emissions depend on average power draw and carbon-intensity assumptions, which can vary across blocked comparisons due to transient hardware allocation and background load. This reinforces the case for reporting both runtime and CO$_2$e, and for interpreting absolute grams as conditional on execution conditions while emphasising relative within-block differences. In addition, it clarifies why runtime remains a valuable benchmark outcome. Runtime is transparent and reproducible, while CO$_2$e provides an environmentally interpretable metric that captures where computation concentrates \citep{strubell2019energy,henderson2020reporting}.

Taken together, the evidence supports a balanced view. LLM-based coding assistance can meaningfully increase research productivity in economics, including for survey-style workflows, but efficiency is not automatic. The key lever is not a new algorithm, but a clearer articulation of what is worth computing. For that reason, our results motivate a lightweight set of practices that researchers can adopt without changing substantive research goals. The aim is not to over-interpret footprint estimates, but to make computational implications visible, comparable, and easier to govern when surveys rely on AI-assisted pipelines.

More broadly, a small set of operational practices can improve computational governance in GenAI-assisted empirical and survey workflows without changing substantive research goals. In practice, this involves consolidating configuration to avoid hidden recomputation (for example, fixing software versions, seeds, and preprocessing choices, and caching intermediate artefacts), constraining exploration explicitly (for example, by using coarse-to-fine searches, pre-defined stopping rules, and decision-focused diagnostics that can be encoded directly in prompts), and computing intensive post-estimation outputs only when they are decision-relevant rather than for every intermediate candidate.

To make evidence more comparable across studies, computational workflows that rely on AI-assisted pipelines can also report a minimal set of implementation details:
\begin{itemize}
\item \textit{Compute context.} Clarify whether execution is local or cloud-based and, where available, report the broad hardware class and key software versions.
\item \textit{Workload descriptors.} Summarise scale and intensity, such as corpus size, preprocessing choices, and the number of candidate specifications or full model fits executed.
\item \textit{Footprint metrics.} Report runtime alongside an energy or emissions estimate, and state the main assumptions when estimates are model-based.
\end{itemize}

These disclosures are intentionally modest. Their purpose is to support basic computational governance as GenAI lowers the cost of iteration, and to reduce the risk that scaling research production mechanically implies scaling avoidable computation.
\clearpage

\bibliographystyle{bde-en}
\bibliography{bibliografia}

\clearpage
\appendix

\section{Appendix}
\subsection{Green AI Literature Mapping Details}\label{sec:lda_method}

This appendix provides methodological details and supplementary material supporting the Green AI literature mapping presented in Section~\ref{sec:greenai_map}. The purpose is twofold. First, we document the construction of the Green AI corpus and the procedures used to isolate contributions that explicitly study the environmental footprint of AI systems. Second, we report additional results and diagnostics from the topic-model-based mapping that underpin the interpretation presented in the main text.

The Green AI corpus is constructed from arXiv preprints retrieved via keyword- and category-based queries. arXiv is chosen as the primary source because Green AI is a recent and technically specialised research area, with a large share of contributions disseminated first as working papers in computer science venues. Restricting the corpus to peer-reviewed outlets would mechanically under-represent current research activity in a fast-moving field. Moreover, arXiv is particularly representative of dissemination norms in computer science, where methodological and systems-oriented AI research is typically circulated as preprints before journal publication. These features make arXiv a natural and sufficiently broad source for mapping the Green AI literature.

\begin{figure}[H]
    \centering
    \includegraphics[width=0.95\textwidth]{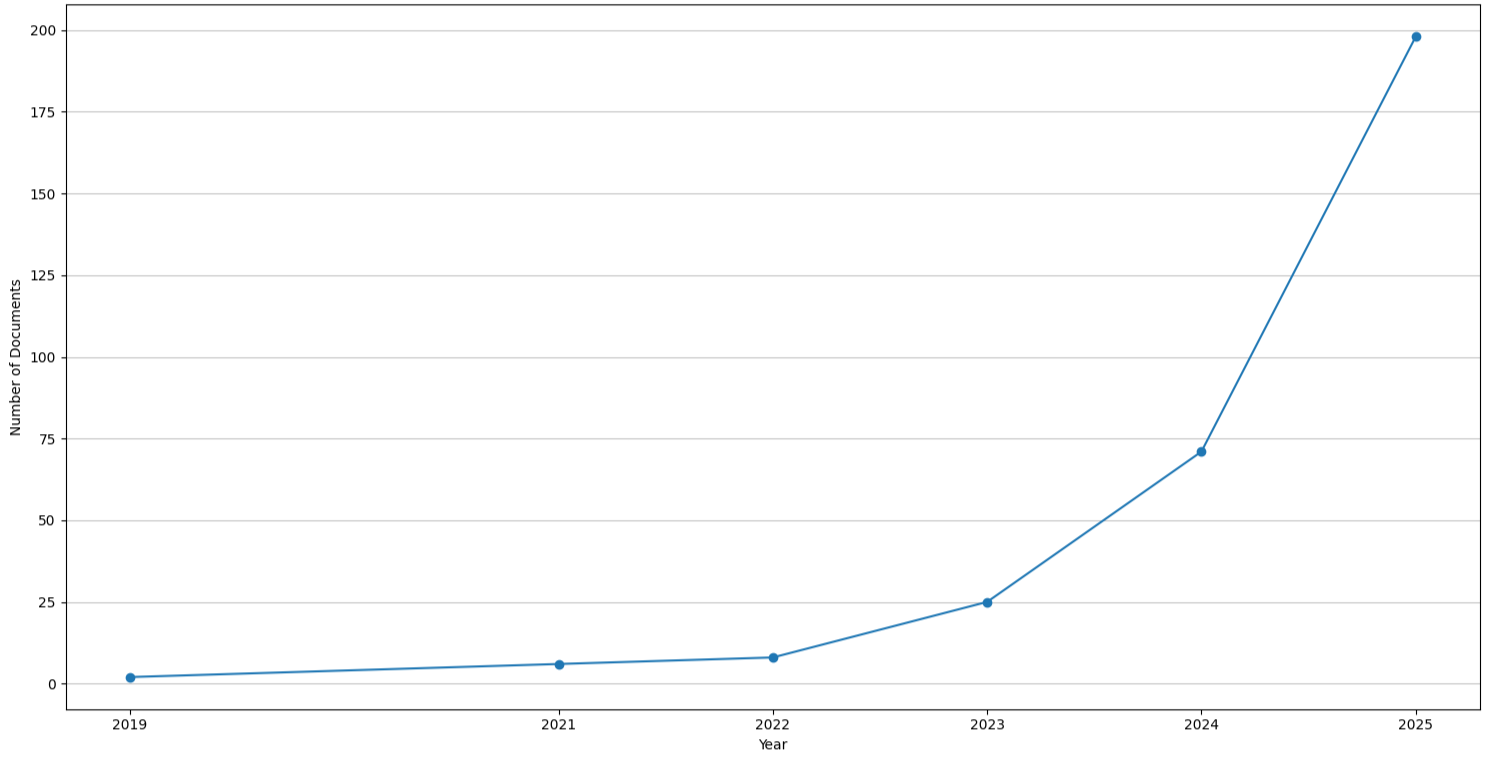}
    \caption{Yearly evolution of the retained Green AI corpus. Bars report the number of arXiv records classified as Green AI (titles and abstracts, after filtering) by publication year, illustrating the recent acceleration in relevant preprints.}
    \label{fig:corpus_evolution}
\end{figure}

To make this motivation more transparent, Figure~\ref{fig:corpus_evolution} reports the yearly evolution of the number of Green AI-relevant documents retained in our final corpus. The sharp increase in recent years supports the choice of arXiv as a primary data source for mapping a fast-moving, technically specialised literature.

To distinguish Green AI papers from applications of AI to environmental or sustainability problems (AI for Green / AI for Environment), we apply a large language model (ChatGPT 5.2) classifier to titles and abstracts using a fixed prompt and explicit inclusion rules.\footnote{To mitigate concerns about prompt-induced endogeneity, we conducted a small manual validation exercise. One author reviewed a subset of 50 retained documents (titles and abstracts) and confirmed that their primary contribution matches the Green AI inclusion rule (environmental footprint/efficiency of AI systems).} The classifier assigns each paper to exactly one category based on its primary research contribution. The full prompt used for this classification step is reported below to ensure transparency and reproducibility.

\begin{tcolorbox}[
  colback=white,
  colframe=black,
  boxrule=0.4pt,
  width=\linewidth,
  sharp corners,
  left=6pt,
  right=6pt,
  top=6pt,
  bottom=6pt,
  breakable
]
\small
\ttfamily
You are assisting in the construction of a research corpus on Green AI. You will be provided with the title and abstract of academic papers retrieved from arXiv. Your task is to classify each paper into exactly one of the following categories, based on its primary research contribution.

\medskip
\textbf{Categories:}

\smallskip
1. \textbf{Green AI}\\
Papers whose main research question explicitly concerns the environmental impact, energy consumption, computational efficiency, or carbon footprint of artificial intelligence systems themselves.

\medskip
2. \textbf{AI for Green / AI for Environment}\\
Papers that apply AI methods to environmental, climate, energy, sustainability, or green finance problems, but do not analyse the environmental footprint or efficiency of AI systems themselves.

\medskip
3. \textbf{Other / Not relevant}\\
Papers that do not clearly fall into either of the two categories above.

\medskip
\textbf{Classification rules:}

\smallskip
- Base the classification strictly on the main contribution described in the title and abstract.\\
- Classify a paper as Green AI only if the environmental cost or efficiency of AI systems is a central object of study.\\
- If the contribution is ambiguous, select the closest category based on the dominant focus of the abstract.

\medskip
For each paper, return the category label and a one-sentence justification.
\end{tcolorbox}

Abstracts retained as Green AI are preprocessed using standard deterministic steps, including lowercasing, tokenisation, English stopword removal, and lemmatisation. Topic modelling is performed using Latent Dirichlet Allocation (LDA) \citep{blei2003lda}, an unsupervised generative model in which each document is represented as a mixture of latent topics and each topic is characterised by a probability distribution over words. LDA is estimated using variational inference as implemented in \texttt{gensim}. Estimated topic--word probabilities $p(w \mid k)$ form the basis for topic interpretation and labeling.

Topic models are estimated for $K \in \{5,\dots,15\}$ topics. Model selection combines quantitative diagnostics with qualitative interpretability. As reported in the main text, coherence improves markedly up to $K=7$ and exhibits diminishing gains thereafter, while higher-$K$ solutions become less semantically distinct upon inspection. These considerations motivate the parsimonious $K=7$ specification used throughout the paper. The $K=7$ solution is also stable under qualitative inspection and exploratory topic-interpretability checks.

Because LDA is an unsupervised method, numeric topic identifiers are arbitrary and may differ across outputs. To ensure consistent presentation across tables and figures, we align internal topic IDs produced by \texttt{gensim} with the presentation topics used in the paper by matching their highest-probability tokens. Table~\ref{tab:topic_id_mapping} reports this mapping, which is used consistently throughout Section~\ref{sec:greenai_map} and the associated figures.

\begin{table}[H]
\centering
\caption{Mapping between internal LDA topic IDs (0--6) and presentation topics (1--7).}
\label{tab:topic_id_mapping}
\begin{tabular}{@{}c c p{0.62\linewidth}@{}}
\toprule
Internal ID & Presentation Topic & Matching cues (high-probability tokens) \\
\midrule
0 & 1 & model, training, carbon, footprint, performance \\
1 & 2 & energy, system, network, efficiency, consumption \\
2 & 3 & framework, research, paper, sustainable, ai \\
3 & 4 & energy, consumption, environmental, method, efficiency \\
4 & 5 & inference, deep, accuracy, computational, task \\
5 & 6 & task, learning, efficiency, computational, performance \\
6 & 7 & attack, algorithm, power, carbon, efficiency \\
\bottomrule
\end{tabular}
\end{table}

\begin{figure}[H]
\centering
\caption{Word clouds for the $K=7$ Green AI topics. Word size is proportional to the estimated topic--word probability $p(w\mid k)$ from the LDA model fitted on preprocessed abstracts (lowercased, stopwords removed, lemmatised). Topic numbering follows the presentation mapping in Table~\ref{tab:topic_id_mapping}.}
\label{fig:wordclouds_topics}
\begin{subfigure}[t]{0.32\textwidth}
  \centering
  \includegraphics[width=\linewidth]{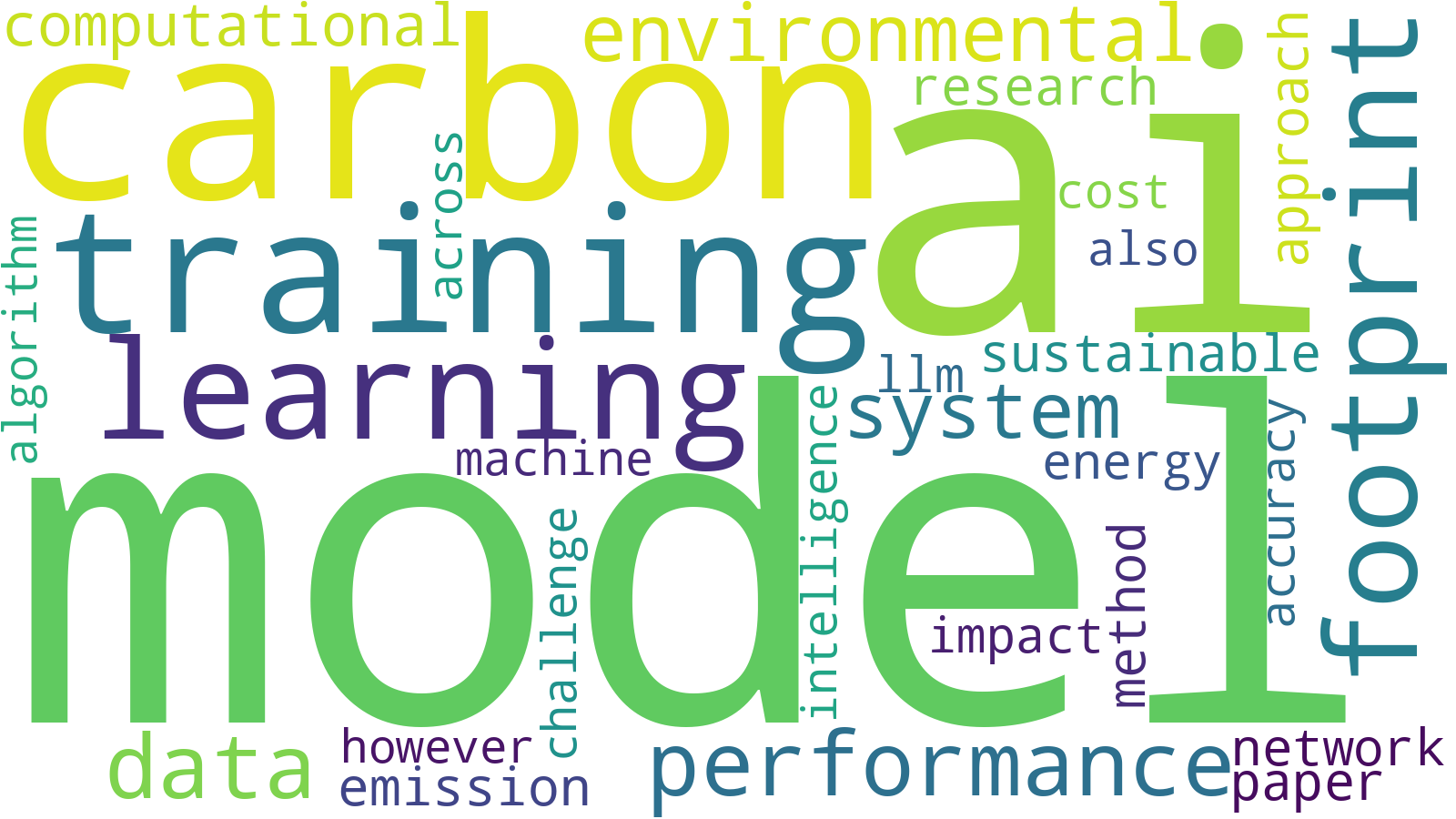}
  \caption{Topic 1: Training footprint}
\end{subfigure}
\begin{subfigure}[t]{0.32\textwidth}
  \centering
  \includegraphics[width=\linewidth]{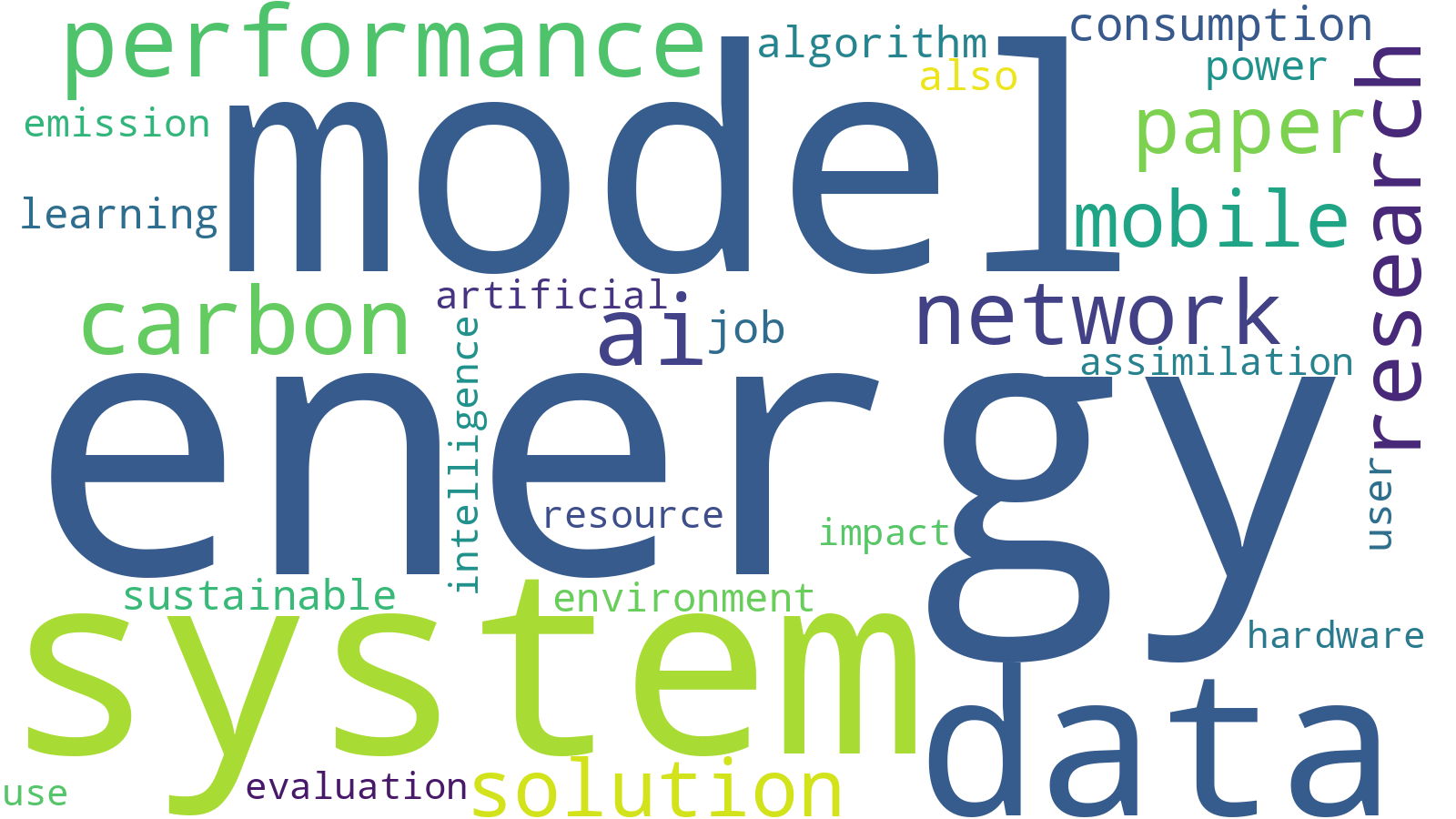}
  \caption{Topic 2: System energy}
\end{subfigure}
\begin{subfigure}[t]{0.32\textwidth}
  \centering
  \includegraphics[width=\linewidth]{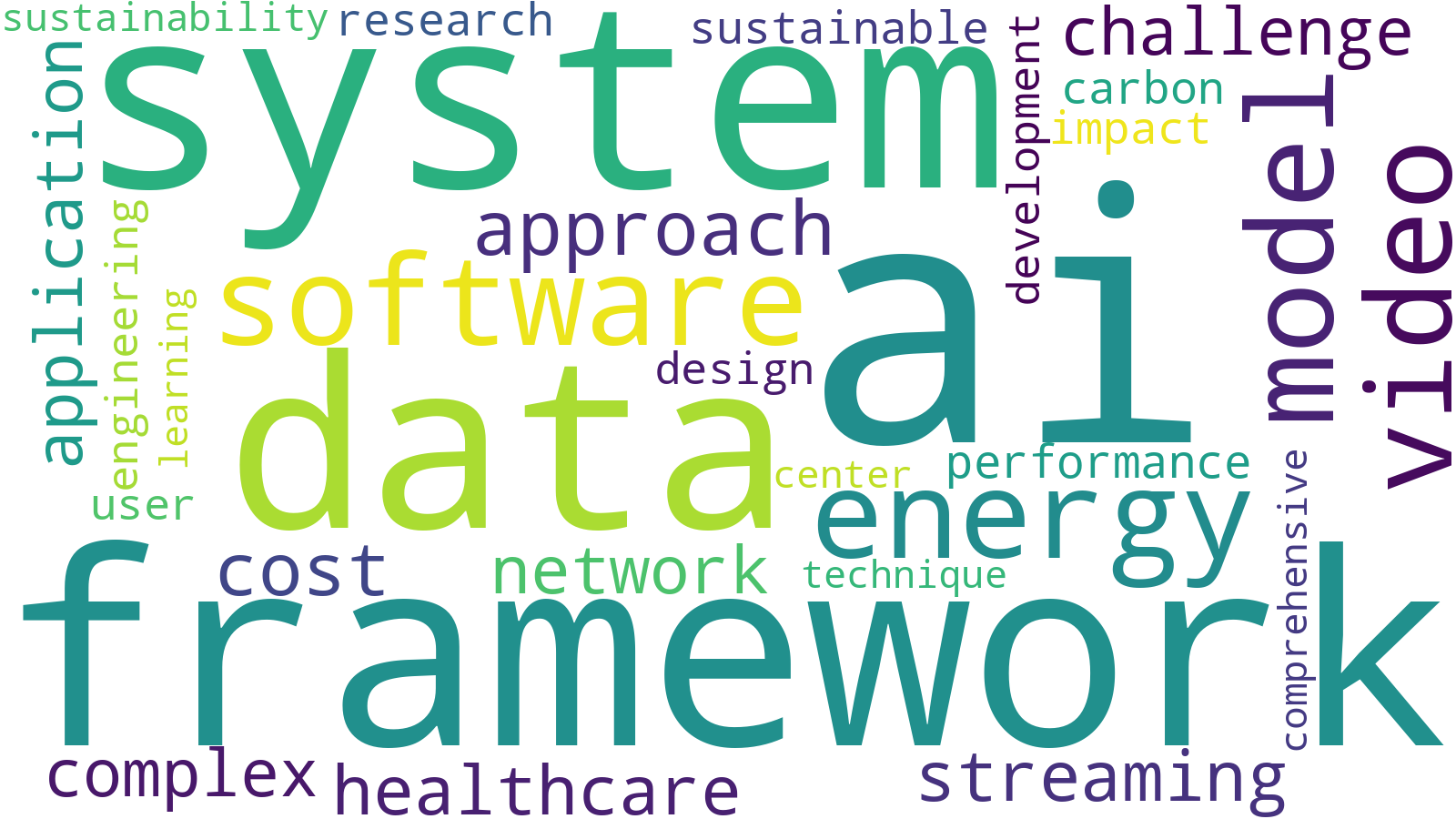}
  \caption{Topic 3: Frameworks}
\end{subfigure}

\begin{subfigure}[t]{0.32\textwidth}
  \centering
  \includegraphics[width=\linewidth]{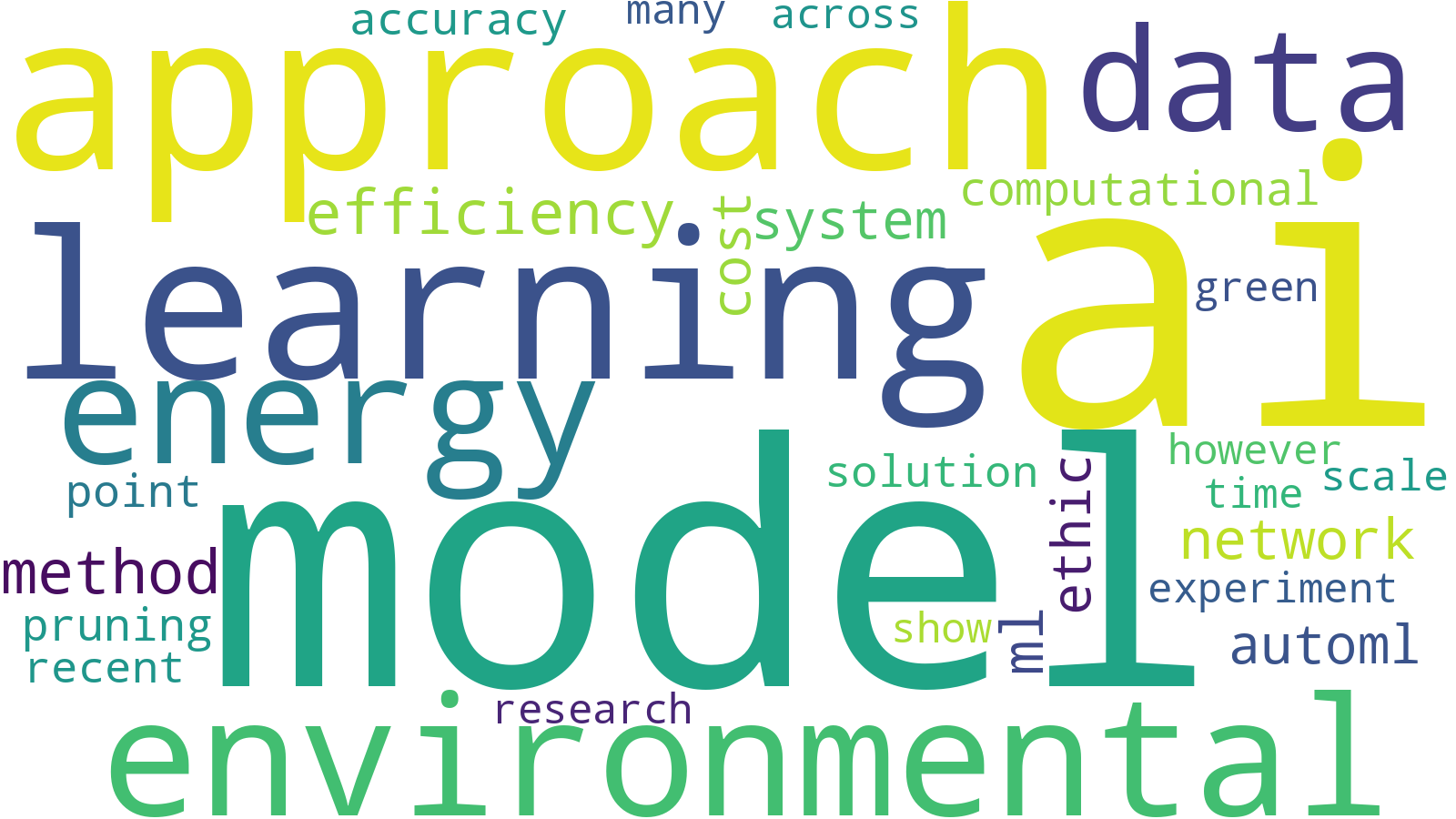}
  \caption{Topic 4: Energy methods}
\end{subfigure}
\begin{subfigure}[t]{0.32\textwidth}
  \centering
  \includegraphics[width=\linewidth]{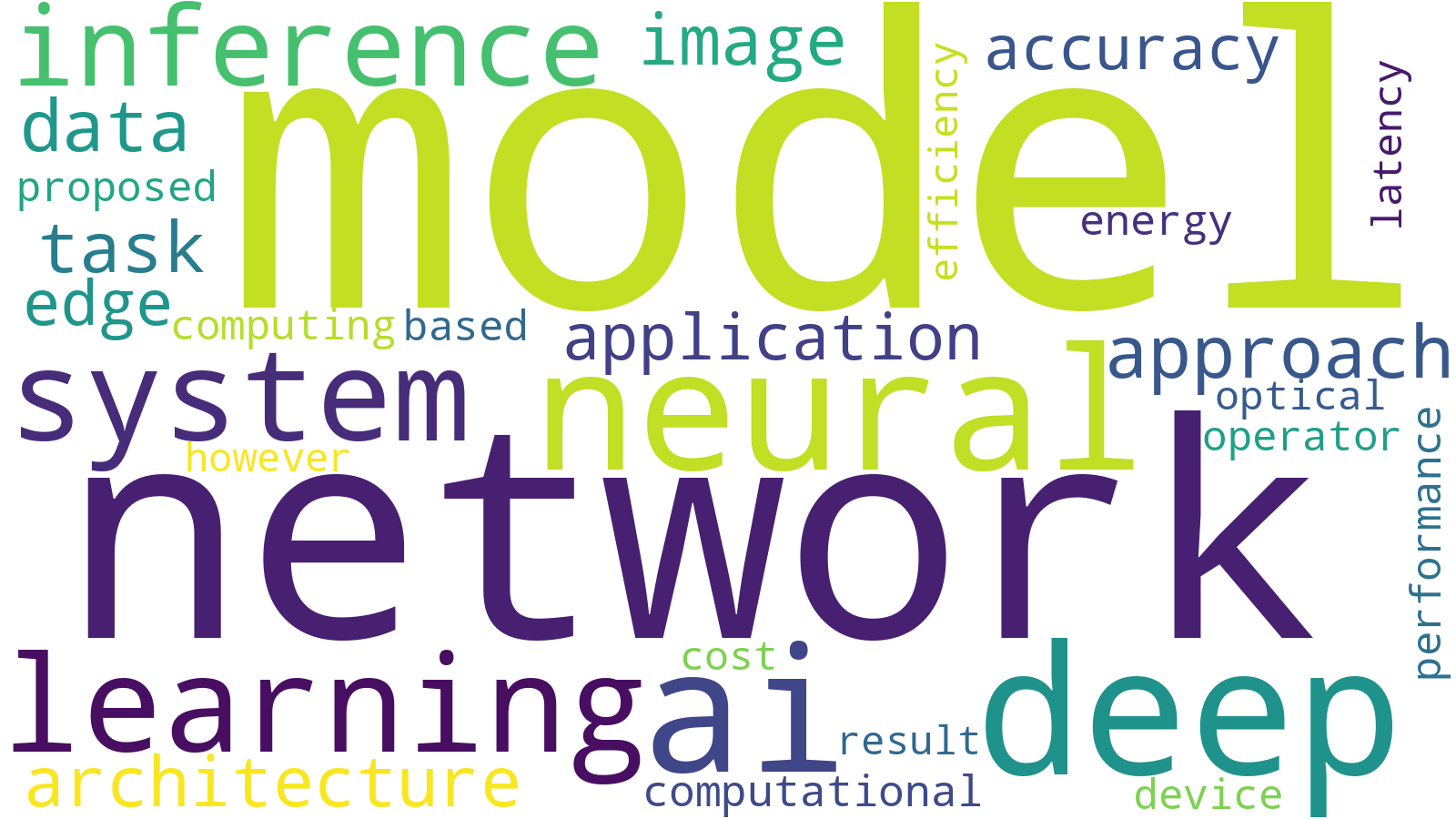}
  \caption{Topic 5: Inference}
\end{subfigure}
\begin{subfigure}[t]{0.32\textwidth}
  \centering
  \includegraphics[width=\linewidth]{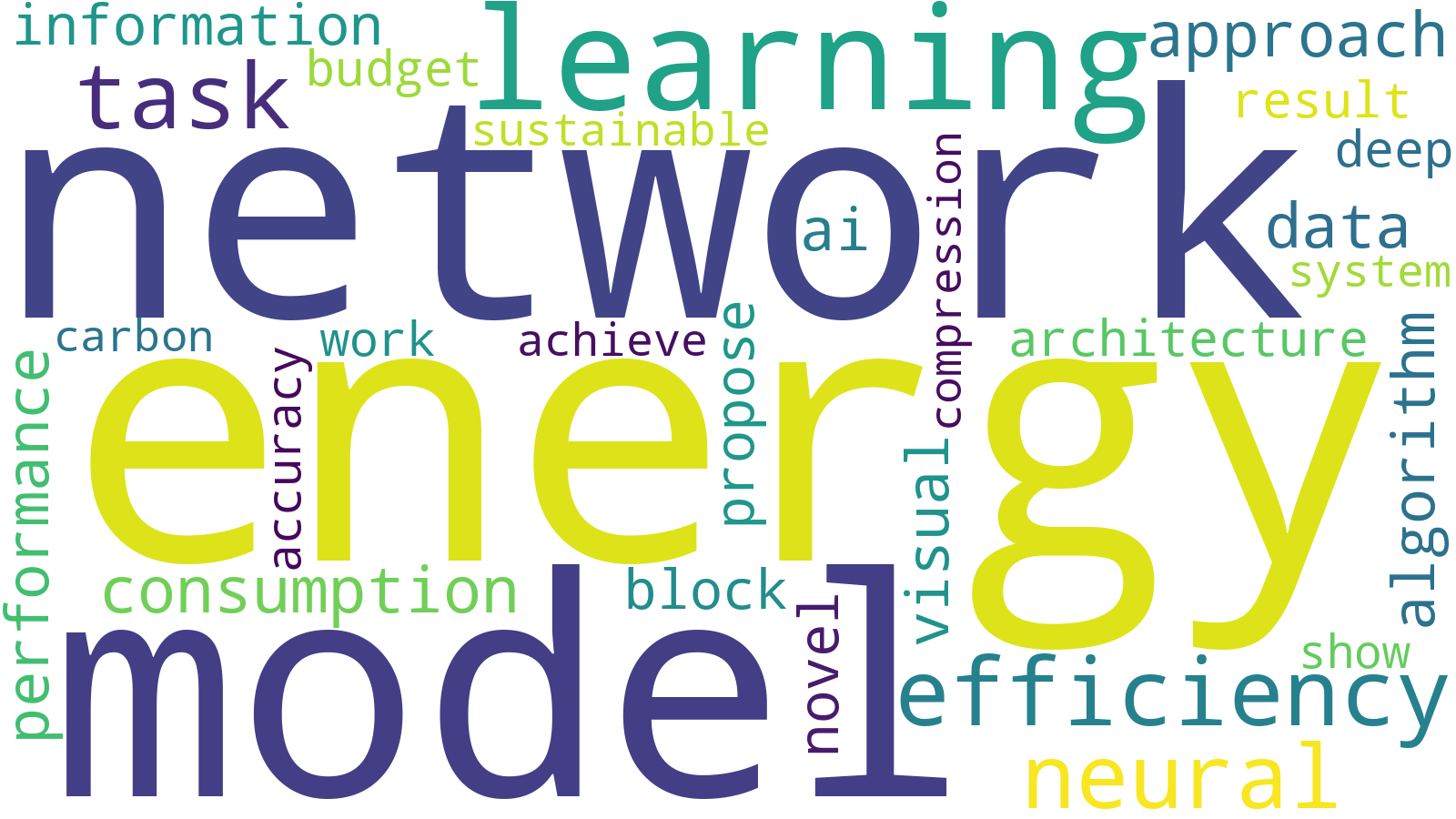}
  \caption{Topic 6: Green algorithms}
\end{subfigure}

\begin{subfigure}[t]{0.32\textwidth}
  \centering
  \includegraphics[width=\linewidth]{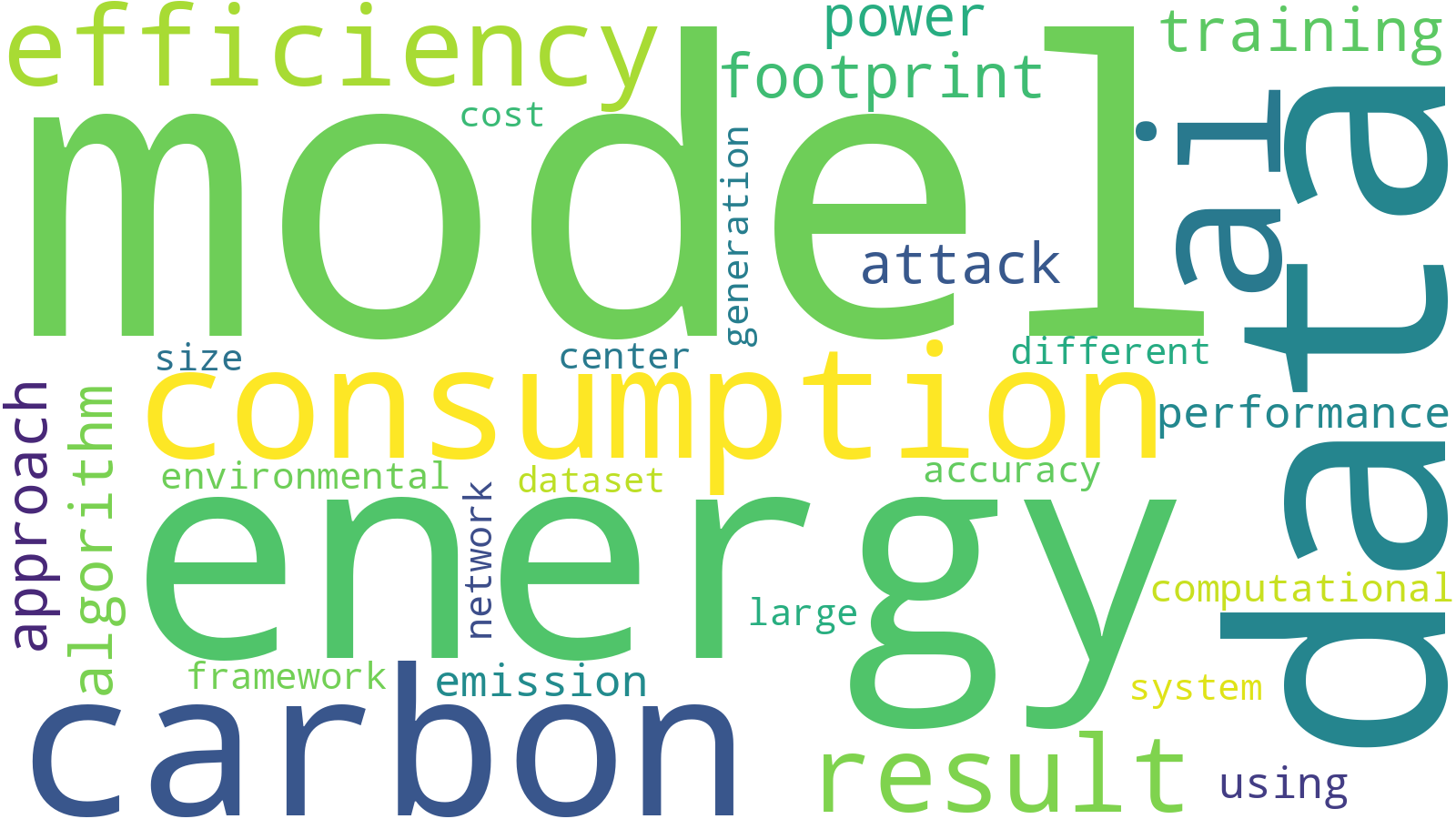}
  \caption{Topic 7: Carbon-aware \& security}
\end{subfigure}
\end{figure}

Table~\ref{tab:topic_tokens} reports the top tokens and their estimated probabilities for each topic in the final $K=7$ model. These distributions underpin the descriptive labels used in the main text and are complemented by the word clouds reported in Figure~\ref{fig:wordclouds_topics}.

\begin{table}[H]
\centering
\caption{Top tokens per topic and estimated probabilities $p(w \mid k)$ in the final $K=7$ model. Topics are ordered according to the presentation mapping in Table~\ref{tab:topic_id_mapping}.}
\label{tab:topic_tokens}
\renewcommand{\arraystretch}{1.15}
\begin{tabular}{c p{0.72\linewidth}}
\toprule
Topic & Top tokens with probabilities \\
\midrule
1 &
0.024 \texttt{model},
0.016 \texttt{ai},
0.010 \texttt{training},
0.010 \texttt{carbon},
0.009 \texttt{learning},
0.007 \texttt{system},
0.007 \texttt{performance},
0.007 \texttt{footprint},
0.006 \texttt{data},
0.006 \texttt{computational}
\\[3pt]

2 &
0.015 \texttt{energy},
0.011 \texttt{system},
0.010 \texttt{ai},
0.008 \texttt{model},
0.007 \texttt{performance},
0.007 \texttt{network},
0.006 \texttt{efficiency},
0.006 \texttt{consumption},
0.006 \texttt{data},
0.005 \texttt{learning}
\\[3pt]

3 &
0.017 \texttt{ai},
0.009 \texttt{data},
0.008 \texttt{energy},
0.006 \texttt{system},
0.006 \texttt{network},
0.006 \texttt{framework},
0.005 \texttt{research},
0.005 \texttt{paper},
0.005 \texttt{approach},
0.005 \texttt{sustainable}
\\[3pt]

4 &
0.019 \texttt{ai},
0.017 \texttt{model},
0.013 \texttt{energy},
0.009 \texttt{data},
0.006 \texttt{efficiency},
0.006 \texttt{consumption},
0.005 \texttt{learning},
0.005 \texttt{environmental},
0.005 \texttt{approach},
0.005 \texttt{method}
\\[3pt]

5 &
0.018 \texttt{model},
0.009 \texttt{network},
0.009 \texttt{inference},
0.008 \texttt{deep},
0.007 \texttt{learning},
0.007 \texttt{accuracy},
0.007 \texttt{computational},
0.007 \texttt{task},
0.007 \texttt{energy},
0.006 \texttt{performance}
\\[3pt]

6 &
0.014 \texttt{energy},
0.012 \texttt{model},
0.007 \texttt{efficiency},
0.007 \texttt{learning},
0.007 \texttt{task},
0.005 \texttt{network},
0.005 \texttt{computational},
0.005 \texttt{performance},
0.005 \texttt{consumption},
0.004 \texttt{accuracy}
\\[3pt]

7 &
0.019 \texttt{model},
0.008 \texttt{energy},
0.008 \texttt{data},
0.007 \texttt{ai},
0.007 \texttt{efficiency},
0.007 \texttt{attack},
0.006 \texttt{carbon},
0.006 \texttt{power},
0.006 \texttt{algorithm},
0.005 \texttt{result}
\\
\bottomrule
\end{tabular}
\end{table}

\begin{figure}[H]
    \centering
    \includegraphics[width=0.85\textwidth]{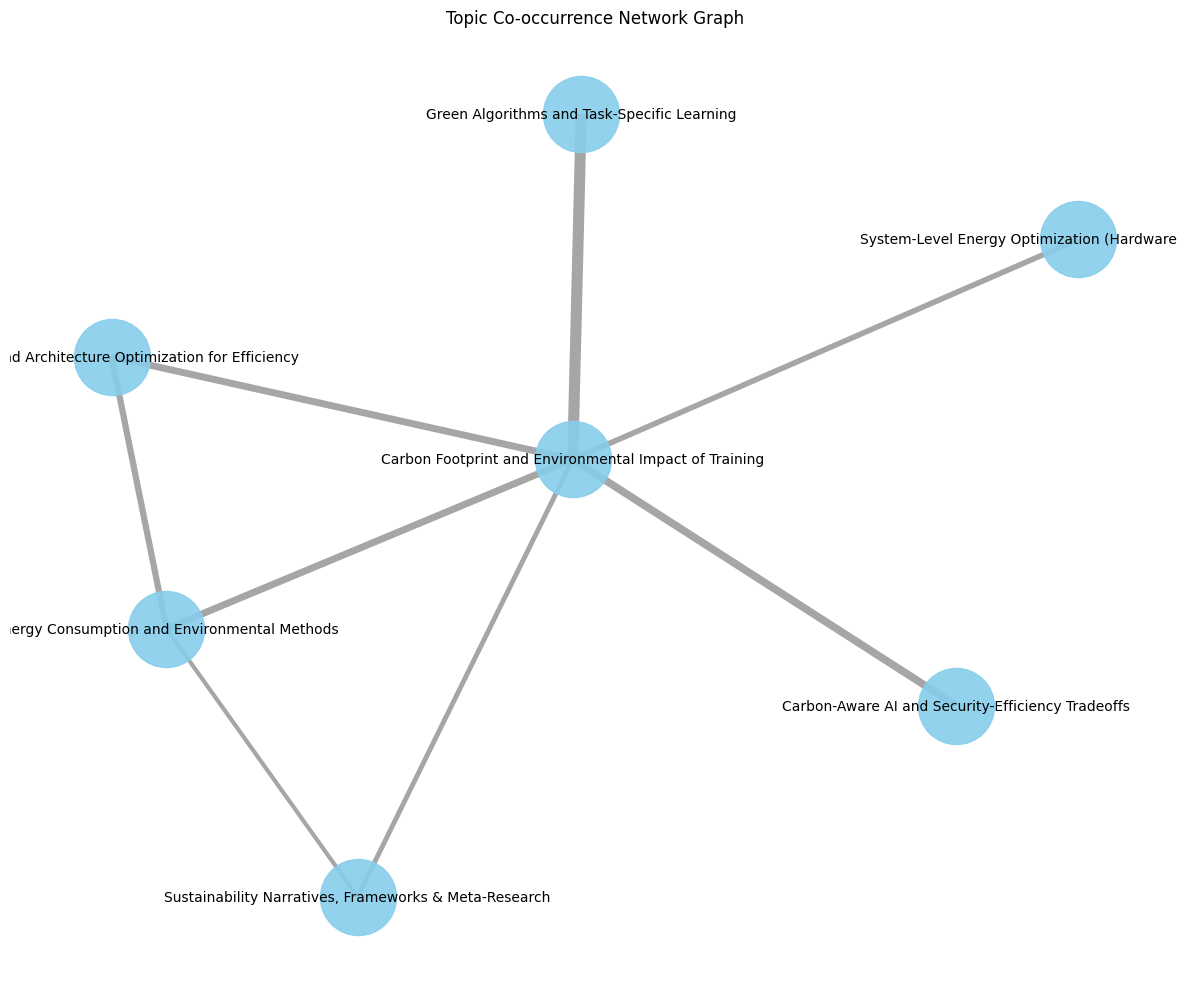}
    \caption{Topic co-occurrence network for the $K=7$ Green AI map. Nodes are topics and edge weights reflect document-level co-occurrence frequency (topics that jointly receive non-trivial weight within the same abstract), offering a graph view of the heatmap in Figure~\ref{fig:coocurrence}.}
    \label{fig:network}
\end{figure}

To characterise how Green AI themes interact, we analyse topic co-occurrence across documents. Figure~\ref{fig:coocurrence} in the main text reports the corresponding heatmap. Figure~\ref{fig:network} provides a network representation of the same structure, where nodes represent topics and edge weights reflect co-occurrence frequency. The network highlights the central role of training-related footprint research and its connections to system-level optimisation and inference efficiency.

Finally, temporal dynamics are examined by tracking the dominant topic assigned to each document over time. Figure~\ref{fig:topic_evolution} in the main text reports the yearly evolution of dominant topic counts, highlighting the rapid expansion of the Green AI literature and the especially strong growth of inference and architecture optimisation contributions in recent years. To complement the token tables and provide an intuitive summary of each theme, we report in Figure~\ref{fig:wordclouds_topics} word clouds for all $K=7$ topics below.

\clearpage

\subsection{Green prompt templates}\label{app:prompts_green}

\paragraph{Green soft prompt (generic energy-efficiency request).}
\begin{tcolorbox}[
  colback=white,
  colframe=black,
  boxrule=0.4pt,
  width=\linewidth,
  sharp corners,
  left=6pt,
  right=6pt,
  top=6pt,
  bottom=6pt,
  breakable
]
\small
\ttfamily
You are generating Python code to run in Google Colab.

\medskip
Assume the following variables already exist in the notebook:

\smallskip
CSV\_PATH (Path)\\
OUTPUT\_DIR (Path)

\medskip
Assume standard data science libraries are available (pandas, numpy, gensim, nltk, matplotlib).

\medskip
Task:

\smallskip
Load the CSV from CSV\_PATH.\\
Use title + abstract as the text (if both are available; otherwise use abstract).\\
Preprocess the data.\\
Explore topic models by training LDA with different numbers of topics from 5 to 15 (inclusive).\\
For each model: print the topics.\\
For each model: compute and report a topic coherence score (c\_v).

\smallskip
For each trained model:\\
\quad Compute the document--topic distributions for all documents.\\
\quad Create a simple time series view showing how the number of documents per topic evolves over time.\\
\quad Produce a table that lists the most relevant documents for each topic.\\
Save the key outputs for each K to OUTPUT\_DIR so I can inspect them later (including the coherence value).

\medskip
Reproducibility:\\
Use fixed random seeds where applicable (e.g., LDA random\_state=42) so results are reproducible.

\medskip
Energy efficiency request:\\
Please write the code with an energy-efficient mindset: avoid wasted work and unnecessary computation where possible, while still completing all requested tasks.

\medskip
The code should run end-to-end in one cell with no manual steps.\\
Return ONLY the Python code for that single cell.
\end{tcolorbox}

\clearpage

\paragraph{Green computational constraints prompt (coarse $K$ grid and implementation guidance).}
\begin{tcolorbox}[
  colback=white,
  colframe=black,
  boxrule=0.4pt,
  width=\linewidth,
  sharp corners,
  left=6pt,
  right=6pt,
  top=6pt,
  bottom=6pt,
  breakable
]
\small
\ttfamily
You are generating Python code to run in Google Colab.

\medskip
Assume the following variables already exist in the notebook:
\smallskip
- CSV\_PATH (Path)\\
- OUTPUT\_DIR (Path)

\medskip
Assume standard data science libraries are available (pandas, numpy, gensim, nltk, matplotlib).

\medskip
Task (same analysis, but implement it efficiently):

\smallskip
1) Load the CSV from CSV\_PATH.\\
2) Use title + abstract as the text (if both are available; otherwise use abstract).\\
3) Preprocess the data (standard deterministic steps).\\
4) Explore topic models by training LDA with different numbers of topics using a coarse pass over K (instead of evaluating every integer K from 5 to 15).\\
\quad - Choose a reasonable coarse grid within [5, 15] that still lets you inspect how topics change with K.\\
\quad - For each trained model: print the topics and compute/report coherence (c\_v).\\
5) For each trained model:\\
\quad - Compute the document--topic distributions for all documents.\\
\quad - Create a time series view of number of documents per topic over time.\\
\quad - Produce a table listing the most relevant documents for each topic.\\
6) Save the key outputs for each evaluated K to OUTPUT\_DIR (including coherence values).

\medskip
Reproducibility:\\
- Use fixed random seeds where applicable (e.g., LDA random\_state=42).

\medskip
Efficiency and implementation guidance (treat as best practices, do not change the analysis logic):\\
- Keep the pipeline minimal and avoid unnecessary work.\\
- Prefer single-pass preprocessing and filtering where possible.\\
- Avoid creating redundant full-text copies and avoid repeated conversions/materializations.\\
- Use lightweight data structures; preallocate NumPy arrays for large numeric outputs (doc-topic matrices).\\
- Avoid expensive per-document Python overhead where you can (vectorize or streamline loops where reasonable).\\
- Do not add extra diagnostics, parameter sweeps beyond the requested K exploration, or additional plots.

\medskip
Preprocessing (standard deterministic steps):\\
- lowercase\\
- remove URLs\\
- remove punctuation and digits\\
- tokenize on whitespace\\
- remove English stopwords (nltk)\\
- lemmatize with WordNetLemmatizer\\
- drop documents with fewer than 5 tokens after preprocessing

\medskip
LDA parameters (fixed for every model):\\
- random\_state = 42\\
- passes = 10\\
- iterations = 200\\
- eval\_every = None

\medskip
The code should run end-to-end in one cell with no manual steps.\\
Return ONLY the Python code for that single cell.
\end{tcolorbox}

\clearpage

\paragraph{Decision-driven stewardship prompt (early stopping and compute only for $K^\ast$).}
\begin{tcolorbox}[
  colback=white,
  colframe=black,
  boxrule=0.4pt,
  width=\linewidth,
  sharp corners,
  left=6pt,
  right=6pt,
  top=6pt,
  bottom=6pt,
  breakable
]
\small
\ttfamily
You are generating Python code to run in Google Colab.

\medskip
Assume the following variables already exist in the notebook:
\smallskip
- CSV\_PATH (Path)\\
- OUTPUT\_DIR (Path)

\medskip
Assume standard data science libraries are available (pandas, numpy, gensim, nltk, matplotlib).

\medskip
Task (same analysis goal, but with decision-focused efficiency):

\smallskip
1) Load the CSV from CSV\_PATH.\\
2) Use title + abstract as the text (if both are available; otherwise use abstract).\\
3) Preprocess the data.\\
4) Select the number of topics K within [5, 15] by balancing insight and computational cost:\\
\quad - Evaluate LDA models sequentially starting from K=5 and increasing K by 2 (i.e., 5, 7, 9, 11, 13, 15).\\
\quad - For each evaluated K:\\
\qquad * Print the topics.\\
\qquad * Compute and report topic coherence (c\_v).\\
\quad - Early stopping rule (economic/decision constraint):\\
\qquad * Track the best coherence seen so far.\\
\qquad * If the absolute improvement in coherence (c\_v) versus the current best is $<0.005$ for two consecutive evaluated K values, stop evaluating further K and keep the best K found so far.\\
\quad - Save a single summary table of evaluated K values and their coherence scores.

\smallskip
5) After choosing the best K:\\
\quad - Compute the document--topic distributions for all documents for that best model only (do not compute doc-topic matrices for every evaluated K).\\
\quad - Create a simple time series view showing how the number of documents per (dominant) topic evolves over time (for the chosen K only).\\
\quad - Produce a table listing the most relevant documents for each topic, but keep it decision-focused:\\
\qquad * Only keep the top 3 documents per topic (ranked by topic probability).

\smallskip
6) Save the key outputs for the chosen best K to OUTPUT\_DIR, plus the K-selection summary:\\
\quad - topics table (top words/weights per topic)\\
\quad - document-topic matrix for the chosen K\\
\quad - topic counts by year (time series table)\\
\quad - top documents by topic (top 3 per topic)\\
\quad - dictionary and trained LDA model\\
\quad - coherence / K-selection summary table (all evaluated K + coherence + stop reason if stopped early)

\medskip
Reproducibility:\\
- Use fixed random seeds where applicable (LDA random\_state=42).

\medskip
Preprocessing (standard deterministic steps):\\
- lowercase\\
- remove URLs\\
- remove punctuation and digits\\
- tokenize on whitespace\\
- remove English stopwords (nltk)\\
- lemmatize with WordNetLemmatizer\\
- drop documents with fewer than 5 tokens after preprocessing

\medskip
LDA parameters (fixed for every evaluated model):\\
- random\_state = 42\\
- passes = 10\\
- iterations = 200\\
- eval\_every = None

\medskip
The code should run end-to-end in one cell with no manual steps.\\
Return ONLY the Python code for that single cell.
\end{tcolorbox}

\clearpage

\subsection{Paired $t$-tests for runtime and emissions}\label{app:ttests}

To assess whether each prompting strategy produces statistically detectable changes in computational footprint relative to the naive baseline, we conduct paired $t$-tests across the $n=5$ counterbalanced repetitions. For each repetition $i$, we compute the paired difference in outcomes (naive minus strategy),
\[
\Delta_i = Y^{\text{naive}}_i - Y^{\text{strategy}}_i,
\]
and test $H_0: \mathbb{E}[\Delta]=0$ using
\[
t = \frac{\overline{\Delta}}{s_{\Delta}/\sqrt{n}}, \qquad df=n-1,
\]
where $\overline{\Delta}$ and $s_{\Delta}$ denote the sample mean and standard deviation of the paired differences. Two-sided $p$-values are reported. Emissions are expressed in grams of CO$_2$e for readability.

\begin{table}[H]
\centering
\caption{Paired $t$-tests relative to the naive baseline ($n=5$, $df=4$): runtime differences (seconds).}
\label{tab:ttests_runtime}
\scriptsize
\setlength{\tabcolsep}{5pt}
\renewcommand{\arraystretch}{1.12}
\begin{tabular}{l r r r l}
\toprule
Strategy vs.\ naive & $\overline{\Delta}_T$ & $s_{\Delta_T}$ & $t$ & $p$ \\
\midrule
Green soft & 1.03 & 28.83 & 0.08 & 0.940 \\
Green computational constraints & 1011.07 & 8.11 & 278.85 & $<0.001$ \\
Decision-driven stewardship & 1251.39 & 50.00 & 55.97 & $<0.001$ \\
\bottomrule
\end{tabular}
\normalsize
\end{table}

\begin{table}[H]
\centering
\caption{Paired $t$-tests relative to the naive baseline ($n=5$, $df=4$): emissions differences (grams CO$_2$e).}
\label{tab:ttests_emissions}
\scriptsize
\setlength{\tabcolsep}{5pt}
\renewcommand{\arraystretch}{1.12}
\begin{tabular}{l r r r l}
\toprule
Strategy vs.\ naive & $\overline{\Delta}_E$ & $s_{\Delta_E}$ & $t$ & $p$ \\
\midrule
Green soft & 0.010 & 0.063 & 0.36 & 0.734 \\
Green computational constraints & 5.149 & 0.041 & 278.81 & $<0.001$ \\
Decision-driven stewardship & 2.532 & 0.101 & 56.04 & $<0.001$ \\
\bottomrule
\end{tabular}
\normalsize
\end{table}

\subsection{Robustness checks for small-sample paired comparisons}\label{app:robustness_paired}

Because each blocked comparison includes $n=5$ paired repetitions, we complement paired $t$-tests with two non-parametric diagnostics. First, we report percentile bootstrap 95\% confidence intervals for the mean and the median paired savings, obtained by resampling the $n=5$ paired differences with replacement (percentile bootstrap, $B=20{,}000$ resamples). Second, we report an exact two-sided sign test based on the number of positive paired savings. The sign test is intentionally conservative at small $n$: even when all five paired savings are positive, the exact two-sided $p$-value equals $0.0625$. We therefore use it as a directional robustness check rather than as a substitute for effect-size and interval reporting.

\begin{table}[H]
\centering
\caption{Robustness checks for paired runtime savings relative to the within-block naive baseline ($n=5$ paired repetitions). We report mean and median paired savings, non-parametric bootstrap 95\% confidence intervals (percentile, $B=20{,}000$ resamples), and an exact two-sided sign test based on the number of positive paired savings.}
\label{tab:robustness_runtime}
\scriptsize
\setlength{\tabcolsep}{3pt}
\renewcommand{\arraystretch}{1.12}
\begin{tabular}{@{}l r r p{2.55cm} p{2.55cm} c r@{}}
\toprule
Strategy & Mean & Median & \makecell[l]{Boot CI95\\(mean)} & \makecell[l]{Boot CI95\\(median)} & \makecell[c]{Pos.} & \makecell[r]{$p$} \\
\midrule
Soft & 1.03 & 2.98 & [-23.83, 20.40] & [-44.99, 31.33] & 3/5 & 1.0000 \\
Comp.\ constraints & 1011.07 & 1011.16 & [1004.62, 1017.11] & [999.72, 1020.54] & 5/5 & 0.0625 \\
Decision-driven & 1251.39 & 1244.57 & [1214.45, 1292.22] & [1197.23, 1323.99] & 5/5 & 0.0625 \\
\bottomrule
\end{tabular}
\normalsize
\end{table}

\begin{table}[H]
\centering
\caption{Robustness checks for paired emissions savings relative to the within-block naive baseline ($n=5$ paired repetitions). We report mean and median paired savings, non-parametric bootstrap 95\% confidence intervals (percentile, $B=20{,}000$ resamples), and an exact two-sided sign test based on the number of positive paired savings. Emissions are expressed in grams of CO$_2$e.}
\label{tab:robustness_emissions}
\scriptsize
\setlength{\tabcolsep}{3pt}
\renewcommand{\arraystretch}{1.12}
\begin{tabular}{@{}l r r p{2.55cm} p{2.55cm} c r@{}}
\toprule
Strategy & Mean & Median & \makecell[l]{Boot CI95\\(mean)} & \makecell[l]{Boot CI95\\(median)} & \makecell[c]{Pos.} & \makecell[r]{$p$} \\
\midrule
Soft & 0.010 & 0.007 & [-0.040, 0.055] & [-0.086, 0.071] & 3/5 & 1.0000 \\
Comp.\ constraints & 5.149 & 5.150 & [5.116, 5.180] & [5.091, 5.197] & 5/5 & 0.0625 \\
Decision-driven & 2.532 & 2.518 & [2.457, 2.615] & [2.423, 2.679] & 5/5 & 0.0625 \\
\bottomrule
\end{tabular}
\normalsize
\end{table}

\end{document}